\documentclass[a4paper,twocolumn,english,superscriptaddress,accepted=2023-06-07]{quantumarticle}
\pdfoutput=1

\usepackage{hyperref}
\usepackage{forest}
\usetikzlibrary{positioning, quotes}
\usepackage[english]{babel}
\usepackage[utf8]{inputenc}
\usepackage{mathtools,stmaryrd}
\usepackage{esint}
\usepackage{bbm}
\usepackage{esvect}
\usepackage{commath}
\usepackage{tikz}
\usepackage{amsmath}
\usepackage{amssymb}
\usepackage[]{graphicx}
\usepackage{float}
\usepackage[colorinlistoftodos]{todonotes}
\usepackage{changepage}
\usepackage[makeroom]{cancel}
\usepackage{indentfirst}
\usepackage{textcomp}
\usepackage{babel}
\usepackage[numbers,sort&compress]{natbib}
\usepackage{bbm}
\usepackage[normalem]{ulem}
\usepackage[linesnumbered,ruled]{algorithm2e}
\usepackage{braket}

\DeclareRobustCommand{\rchi}{{\mathpalette\irchi\relax}}
\newcommand{\irchi}[2]{\raisebox{\depth}{$#1\chi$}} % used by \rchi

\DeclareMathOperator*{\argmin}{arg\,min}

\begin{document}
\title{Real-Time Krylov Theory for Quantum Computing Algorithms}

\author{Yizhi Shen}
\email{yizhis@lbl.gov}
\affiliation{NASA Ames Research Center, Moffett Field, CA 94035, USA}
\affiliation{KBR, 601 Jefferson St., Houston, TX 77002}
\affiliation{Department of Chemistry, Massachusetts Institute of Technology, Cambridge, Massachusetts 02139, USA}

\author{Katherine Klymko}
\affiliation{NERSC, Lawrence Berkeley National Laboratory, Berkeley, California 94720, USA}

\author{James Sud}
\affiliation{NASA Ames Research Center, Moffett Field, CA 94035, USA}
\affiliation{USRA Research Institute for Advanced Computer Science, Mountain View, CA 94043, USA}

\author{David B. Williams--Young}
\affiliation{Applied Mathematics and Computational Research Division, Lawrence Berkeley National Laboratory, Berkeley, California 94720, USA}

\author{Wibe A. de Jong}
\affiliation{Applied Mathematics and Computational Research Division, Lawrence Berkeley National Laboratory, Berkeley, California 94720, USA}

\author{Norm M. Tubman}
\email{norman.m.tubman@nasa.gov}
\affiliation{NASA Ames Research Center, Moffett Field, CA 94035, USA}
% \date{\today}

\begin{abstract}
Quantum computers provide alternative avenues to access ground and excited state properties of systems difficult to simulate on classical hardware. New approaches using subspaces generated by real-time evolution have shown efficiency in extracting eigenstate information, but the full capabilities of such approaches are still not understood. 
In recent work, we developed the variational quantum phase estimation (VQPE) method, a compact and efficient real-time algorithm to extract eigenvalues using quantum hardware. 
Here we build on that work by theoretically and numerically exploring a generalized Krylov scheme where the Krylov subspace is constructed through a parametrized real-time evolution, applicable to the VQPE algorithm as well as others. 
We establish an error bound that justifies the fast convergence of our spectral approximation. 
We also derive how the overlap with high energy eigenstates becomes suppressed from real-time subspace diagonalization and we visualize the process that shows the signature phase cancellations at specific eigenenergies. 
We investigate various algorithm implementations and consider performance when stochasticity is added to the target Hamiltonian in the form of spectral statistics. 
To demonstrate the practicality of such real-time evolution methods, we discuss its application to fundamental problems in quantum computation such as electronic structure predictions for strongly correlated systems.
\end{abstract}

\maketitle

\section{Introduction}
Quantum computers offer the promise of improvements over their classical counterparts for tackling a class of problems central in the mathematical and physical sciences by encoding information as quantum many-body states. 
However, given current limitations on the assembly and control of scalable quantum computers, efficient usage of quantum resources for specific tasks~\cite{hybrid_correlated,hybrid_qcontrol,training_qcircuit,Mcclean2017, Arute2019, Mcardle2020,lin2022heisenberg_prx,An2021timedependent,lin2021fastinversion} is considered essential in the noisy intermediate-scale quantum (NISQ) era~\cite{nisq_1, nisq_2, Arute2020b}. 
As one of the most prominent algorithms, quantum phase estimation (QPE)~\cite{nielsen_chuang} resolves the core task of Hamiltonian diagonalization but necessitates a relatively high simulation cost. 
Consequently, approaches relying on variational algorithms~\cite{Peruzzo2014_vqe, wecker2015progress, Sung2020, Huggins2020, Arute2020a, Fedorov2021_vqe} have been pursued, focused on balancing resource allocation. 
They generally do so by preparing and measuring parametrized states on a quantum computer while steering parameter updates through optimization routines on a classical computer. 
This hybridization allows for a speedup of high-dimensional problems on near-term hardware, yet comes with complexities depending on choice of the variational ansatz. 
Fortunately, these additional complexities may be alleviated by clever and flexible ansatz design that fully accommodates the architecture of a given quantum device.~\cite{McClean2016_vqe, Takeshita2020}

Among such hybrid quantum-classical approaches, subspace expansion techniques employing real time quantum dynamics~\cite{Parrish2019a,Urbanek2020,Stair2020,cortes2022} have shown evidence of advantages on near-term hardware. 
One representative approach is the so-called variational quantum phase estimation (VQPE) studied and developed recently~\cite{Klymko2021_VQPE}. 
VQPE shares the merits of variational approaches and bypasses conventional optimization procedures by solving generalized eigenvalue equations with information gathered from real-time evolution~\cite{Park1986,Neuhauser1990,Neuhauser1994,Wall1995}, which is unitary and thereby native to quantum hardware. 
Moreover, real-time evolution with VQPE enables access to the excited state manifold and requires quantum measurements merely linear in the dimension of expansion subspace. 
Because of its compactness, VQPE stands out as a promising algorithm for the NISQ era.

Recent theoretical development~\cite{Klymko2021_VQPE} of real-time evolution highlights the phase cancellation intuition for perfect spectral recovery, where eigenspaces of a target Hamiltonian operator are extracted exactly provided $(i)$ the number of evolution timesteps matches the size of the Hilbert space and $(ii)$ the time-evolved phases satisfy a set of geometrically meaningful sum rules. 
However, fulfillment of such phase conditions is only a serious consideration when the full spectrum of the Hamiltonian is needed. In reality, low energy part of the spectrum often suffices under many circumstances of interest. 
In this regard we present a complementary perspective on VQPE for its main use case, where the number of timesteps is kept significantly smaller than the size of Hilbert space, demonstrating that real-time evolution remains powerful for ground and low-lying excited state recovery. 
For generality, we formalize the real-time approach as a parametrizable variant of the Krylov method~\cite{Lanczos1950,parlett1980_eigvalbound,Meyer1989,Manmana2005,Koch2011} with evolution timestep acting as the hyperparameter. 
We suggest weaker phase cancellation conditions for accurate spectral approximation, and examine the effects of stochasticity on observed convergence. 
To illustrate its appealing practicalities, we also discuss how the real-time Krylov theory can be integrated into quantum computing algorithms.

\subsection{Contributions}

In the following sections, we share four main results for understanding the properties of real-time Krylov method based on the generation of states from Hamiltonian evolution. 
We first demonstrate and visualize the convergence for single-step simulation, and then turn to multi-step simulation (Sections \ref{sec:Single step convergence}-\ref{sec:Multi-step convergence}). Next we provide a proof of the convergence with an increasing number of timesteps (Section \ref{sec:Multi-step convergence}). Finally, we consider and assess an iterative implementation of the method for generating real-time states and show how this can further improve the convergence behaviors (Section \ref{sec:Implementation analysis}).
\bigskip

\section{\label{sec:Krylov overview}Theoretical Overview}
\subsection{Review of the Krylov method}
The Krylov subspace method~\cite{parlett1980_eigvalbound} is a common numerical tool to extract useful spectral information from some operator $\hat{H}$ over a Hilbert space $\mathcal{H}$. The method proves particularly powerful for approximating the extreme ends of the operator spectrum. Here we briefly review how the method works and set up the notational convention for the remaining sections. Throughout this work we assume the operators to be self-adjoint, $\hat{H} = \hat{H}^{\dagger}$.

The Krylov method computes the eigenspaces by compressing the target operator, $\hat{H}$, onto a lower-dimensional subspace known as the Krylov subspace, 
\begin{equation}
    K(\Phi_0; N_T) = {\rm span}\left\{ |\Phi_{j} \rangle =  \hat{H}^{j} |\Phi_0 \rangle: j \leq N_{T} \right\},
\end{equation}
where a number of $N_{T}$ repeated $\hat{H}$-multiplications is applied to an initial vector $|\Phi_0 \rangle$. In language of matrix algebra, diagonalization within the Krylov subspace amounts to solving the eigenvalue problem,
\begin{eqnarray}
    \mathbf{H} \Vec{c}_{n} = E_{\Tilde{n}} \mathbf{S} \Vec{c}_{n}, 
    \label{eq:linsystem}
\end{eqnarray}
where $\mathbf{H}$ and $\mathbf{S}$ represent the target and overlap matrices in the Krylov basis,
\begin{eqnarray}
    \begin{split}
        \hspace{0.6 cm} \mathbf{H}_{ij} &= \langle \Phi_i| \hat{H} | \Phi_j \rangle, \\
        \mathbf{S}_{ij} &= \langle \Phi_i | \Phi_j \rangle,
    \end{split}
\end{eqnarray}
while $\vec{c}_{n}$ give the expansion coefficients of an approximate eigenvector having the eigenvalue $E_{\Tilde{n}}$. In practice, an initial vector $|\Phi_0 \rangle$ can be chosen to make the Krylov vectors all linearly independent.
The Krylov method thereby extracts a subset of the target spectrum by factorizing the reduced matrix $\mathbf{H}$. Its efficiency is manifested especially when $N_T \ll {\rm dim}\mathcal{H}$.

\subsection{Generalized Krylov method from unitary action}
We consider a parametrizable variant of the standard Krylov method, where a series of discrete time values are used to exponentiate $\hat{H}$ and thus constitute the free (hyper)parameters. Specifically,  we allow the following generalized notion of Krylov subspace: for the initial vector $|\Phi_0 \rangle$, we apply unitary evolution of the form,
\begin{equation}
    \hat{U}_j = \exp{( -i \hat{H} t_j)},
\end{equation}
where $0 = t_0 < t_1 < \cdots < t_{N_T} < \infty$ records the timestamps of the evolution and $i^2 = -1$ denotes the imaginary unit. The evolved vectors generate a subspace,
\begin{equation}
    K_{\hat{U}}(\Phi_0; N_T) = {\rm{span}} \left\{ |\Phi_{j} \rangle = \hat{U}_j |\Phi_0 \rangle: j \leq N_T \right\},
\end{equation}
over which we can solve the eigenvalue problem. 
The free hyperparameter in this algorithm, the time grid $\vec{t} = (t_1, \cdots, t_{N_T})$, effectively accommodates the linear independence of $\{ |\Phi_{j} \rangle \}_{j=0}^{N_T}$. For linear time grid $t_{j} = j \Delta t$, VQPE reduces to the standard Krylov subspace method applied to the operator, 
\begin{eqnarray}
    \hat{U}(\Delta t) = \sum_{n=1}^{{\rm dim}\mathcal{H}} \exp{\left( -i E_n \Delta t \right)}| n \rangle \langle n|, 
\end{eqnarray}
where $\{ | n \rangle \}_{n}$ and $\{ E_n \}_{n}$ label the true eigenstates and eigenvalues respectively. Note that $\hat{U}$ clearly shares the same eigenstates with our target operator $\hat{H}$.

\subsection{Real-time Krylov method}
Here we provide a summary of the real-time method by outlining its key algorithmic steps. We will refer to the spectrum of $\hat{H}$ interchangeably as a set of energies from now on (the lowest eigenvalue being the ground state energy, \textit{etc}...). For concreteness, we focus on the task of ground state estimation as described in Alg.~\ref{alg:vqpe} below,
\begin{center}
    \begin{algorithm}
    \KwData{time-ordered grid $\{ t_j \}_{j=0}^{N_T}$ with $t_0 = 0$}
     prepare initial state $|\Phi_0 \rangle$\;
     estimate ground state energy $E_{\rm g} = \langle \Phi_0 | \hat{H} | \Phi_0 \rangle$\;
     $j = 0$\;
     $\varepsilon = \infty$\;
    \While{$\varepsilon > \varepsilon_{\rm tol}$ }{
      $j \leftarrow j+1$\;
      $|\Phi_j \rangle =  \hat{U}_j(t_j) | \Phi_0 \rangle $\;
      $\mathbf{H}_{k\ell} = \langle \Phi_k | \hat{H} | \Phi_\ell \rangle, \mathbf{S}_{k\ell} = \langle \Phi_k | \Phi_\ell \rangle  $\;
      Solve $j \times j$ projected eigenvalue problem $\mathbf{H} \Vec{c}_{n} = E_{\Tilde{n}} \mathbf{S} \Vec{c}_{n}$\;
      $\varepsilon \leftarrow \abs{E_{\rm g} - \min\{ E_{\Tilde{n}} \}_{n \leq j
      }}$\;
      $E_{\rm g} \leftarrow \min\{ E_{\Tilde{n}} \}_{n \leq j}$\;
     }
    \caption{ground state estimate (VQPE)}
    \label{alg:vqpe}
    \end{algorithm}    
\end{center}
\noindent where the  algorithm terminates when the convergence variable $\varepsilon$ computed within an iteration, such as set by the difference in consecutive energy estimates, reaches reasonable tolerance $\varepsilon_{\rm tol}$. The tolerance value is user-defined and depends on the specific problem or target operator of interest. For example a molecular problem typically concerns a chemical accuracy of $\varepsilon_{\rm tol} \approx 10^{-3}$ in energy units of Hartree.
\bigskip

\section{\label{sec:Single step convergence}Why does phase cancellation converge so quickly?}
\subsection{\label{sec:Single step} Understanding phase cancellation}
A main goal of this work is to motivate a simple understanding towards the use of real-time states. We first address why superposing these equi-energy states helps generate the ground state, as previous work~\cite{Klymko2021_VQPE} has demonstrated that the ground-state convergence can be reached with a surprisingly small number of real-time states. Here we exploit eigenfunction expansions to demonstrate the suppression of amplitudes on highly energetic eigenstates. Note that such analysis appears natural for purely projective approaches such as the power method~\cite{PowerMethod} and imaginary time evolution~\cite{Motta2020b,Yeter2020}, where the convergence is shown exponentially fast. In this section, we present a visual representation of the single step solution. Later we provide a proof of the convergence as well as visualization of the multi-step solution.

\begin{figure}[H]
\centering
\includegraphics[width=.425\textwidth]{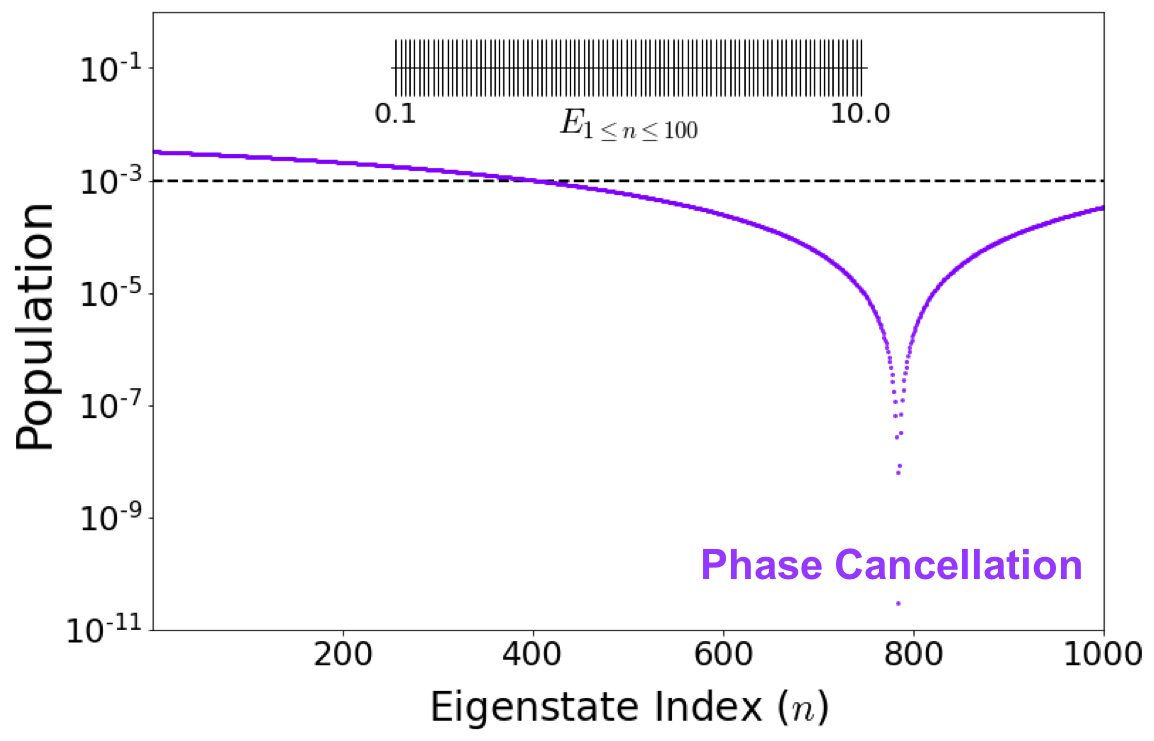}
\caption{Generic single step suppression of a uniformly dense spectrum. Eigenstate population $p_{n} = |\langle n |\Psi_{\rm g} \rangle|^2$ of the approximate ground state $|\Psi_{\rm g} \rangle$ is plotted over the eigenstate index $1 \leq n \leq 1000$. The black dashed curve and solid purple curve show the initial and time-evolved population profiles respectively. The spectral spacing between low-lying eigenstates is also displayed inset for visualization.}
\label{fig:singlestep}
\end{figure}

In Fig.~\ref{fig:singlestep} we demonstrate the single step suppression, a generic behavior that occurs for a dense spectrum across the test simulations. In our demonstration, we start from an equal superposition over the entire Hilbert space and generate our ground state approximation, $|\Psi_{\rm g} \rangle$, by taking a single timestep. The eigenstate amplitudes, $|\langle n |\Psi_{\rm g} \rangle|^2$, exhibit that a single eigenvalue is exactly suppressed, and the nearby spectral region is also suppressed within some width of the minimum.

Recall that we collect two real-time states, namely $|\Phi_0 \rangle$ and $|\Phi_1\rangle = \exp{(-i \hat{H} t_1)}|\Phi_0\rangle$ in a single step. Intuitively, we know from subspace diagonalization that there exists some choice $c_1 \in \mathbb{C}$ satisfying,
\smallskip
\begin{eqnarray}
     \begin{split}
         |\Psi_{\rm g} \rangle &= \frac{|\Phi_0  \rangle +  c_1 | \Phi_1 \rangle}{ \sqrt{1 + \abs{c_1}^2 + 2\Re{(c_1 \mathbf{S}_{01})}}  }, \\
         & \propto \sum_{n=1}^{{\rm dim}\mathcal{H}} z_n | n\rangle + c_1 \sum_{n=1}^{{\rm dim}\mathcal{H}} z_n \exp{(-i\hat{H}t_1)} | n\rangle, \\
         &= \sum_{n=1}^{{\rm dim}\mathcal{H}} z_n \big[ 1 + c_1 \exp{(-i E_n t_1}) \big] |n \rangle,
     \end{split}
\end{eqnarray}

\noindent where $\mathbf{S}_{01} = \langle \Phi_0 | \Phi_1 \rangle$ gives the overlap between states $\ket{\Phi_0}$ and $\ket{\Phi_1}$. In the second line, we make use of an eigenbasis expansion of the real-time states, in terms of the coefficients $z_n = \langle n | \Phi_0 \rangle$, to show the eigenphase evolution. So the amplitude decay at the observed eigenstate arises due to the phase associated with that eigenstate rotated to a value of $-1$, canceling out the $+1$ initial phase along the real axis. Meanwhile the eigenstates nearby are also rotated to fulfill nearly the same phase cancellation, thus there is a finite width to the decay. As eigenstates close in energy will pick up similar phases under real-time evolution, amplitudes on many excited states can be simultaneously suppressed. Accordingly, we expect reduced convergence for much larger timesteps as phases acquired by adjacent eigenstates in the spectrum become more separated.

For concreteness, let us work out the simplest case of time evolution under a single Pauli-$Z$ operator, corresponding physically to Rabi oscillations of spin-1/2 particle in a quantum mechanical description. We denote the $Z$-computational basis by $\ket{\uparrow}$ and $\ket{\downarrow}$ where $Z\ket{\uparrow}=\ket{\uparrow}$ and $Z\ket{\downarrow}=-\ket{\downarrow}$. Starting with the spin polarization 
 $|\Phi_0\rangle = (\ket{\uparrow} + \ket{\downarrow} ) /\sqrt{2}$, we can evolve our initial state by the discrete timestep $t_1 = \pi/2$ so that,
\begin{eqnarray}
    |\Phi_1 \rangle = \frac{ - i (-\ket{\uparrow} + \ket{\downarrow} )}{\sqrt{2}},
    % &=& \frac{\exp{(i\pi/2)}\ket{+} + \exp{(-i\pi/2)} \ket{-}  }{\sqrt{2}},\\
    % &=&
\end{eqnarray}
where the $\ket{\uparrow}$ eigenphase undergoes a sign flip. Phase cancellation in this case is thus explicit: $\ket{\downarrow} = (|\Phi_0\rangle + c_1 |\Phi_1\rangle)/\sqrt{2}$ with $c_1 = i$. We can immediately proceed to the ensuing case of time evolution under a sum of two Pauli-$Z$ operators. For the initial bipartite spin state $\ket{\Phi_0} = (\ket{\uparrow \uparrow} + \ket{\uparrow \downarrow} + \ket{\downarrow \uparrow} + \ket{\downarrow \downarrow})/2$, a linear time grid of $(t_1, t_2) = (\pi/4, \pi/2)$ ensures that $ \ket{\Phi_1} $ and $ \ket{\Phi_2} $ pick up an extra minus sign, respectively, on the $\ket{\uparrow \uparrow}$ and $\{ \ket{\uparrow \downarrow}, \ket{\downarrow \uparrow} \}$ eigenphases relative to the $\ket{\downarrow \downarrow}$ eigenphase. Phase cancellation again facilitates the ground state recovery, $\ket{\downarrow \downarrow} = (\ket{\Phi_0} + c_1 \ket{\Phi_1} + c_2 \ket{\Phi_2})/(1+i)$ with $(c_1, c_2) = (i-1, -i)$. Likewise, taking additional timesteps in a real-time evolution leads to further amplitude suppressions, each peaked around a different eigenstate in the spectrum. This general picture persists beyond simple cases enumerated here and points to the basic phase cancellation heuristics. 
\bigskip

\section{Single step examples and convergence properties}
Now we consider various types of convergence tests tailored for different applications. 
%We remark that VQPE is an advantageous hybrid solver as its implementation on quantum hardware involves relatively shallow circuit. 
In general, the associated quantum circuit enacts 
$(i)$ time evolution of the initial state, $|\Phi_0 \rangle \mapsto \exp{(-i\hat{H} t)}|\Phi_{0} \rangle $, which is difficult to simulate classically, and $(ii)$ subsequent measurements of the matrix elements, $\mathbf{H}_{ij}$ and $\mathbf{S}_{ij}$, for example through the Hadamard test or shadow tomography~\cite{huggins2022unbiasing}. The unitary formulation of VQPE exploits the toeplitz structure of the Hamiltonian and overlap matrices, reducing the number of required measurements to being linear in the number of timesteps. Although we regard the real-time algorithm as broadly suited in many quantum computing applications, we will also discuss scenarios where the algorithm proves inefficient. For example, unstructured search, as discussed in the beginning of the section, turns out to be rather impractical due to implementation barriers which we will describe. 

Let us first focus on the single timestep limit, \textit{i.e.}, $N_T = 1$, such that ${\rm dim}K = 2$. For convenience, we define $Q = {\rm dim}\mathcal{H}$ throughout the remaining sections. 

\subsection{\label{sec:Grover} Unstructured  search}
Given some Boolean function $f: \mathcal{B} \rightarrow \mathbb{Z}_2$ over a set of candidate database elements $\mathcal{B} = \{n\}_{n = 1}^{Q}$, the task of an unstructured search is to locate, without any a priori knowledge of the database structure, the unique flagged element $n_1 \in \mathcal{B}$ for which $f(n_1) = 1$. Such search can be formulated as an eigenspace search through the identification,
\begin{eqnarray}
     \hat{H} = E_1 | 1 \rangle \langle 1 | + E_2 \sum_{n=2}^{Q} | n \rangle \langle n |, ~E_1 < E_2,
     \label{eq:Grover_H}
\end{eqnarray}
where we assume $n_1 = 1$ and $\hat{H}$ acts on the Hilbert space $\mathcal{H} = {\rm span}\left\{ | n \rangle: 1 \leq n \leq Q \right\}$. For any initial state,
\begin{eqnarray}
     |\Phi_0 \rangle = \sum_{n=1}^{Q} z_n |n \rangle,
\end{eqnarray}
a single VQPE step with evolution time $t_1$ will send it to,

\begin{equation}
\begin{split}
|\Phi_1 \rangle = & \exp(-iE_1 t_1) \big[ z_1 |1 \rangle \\
&\hspace{1.5 cm} + \exp(-i \Delta E t_1) \sum_{n=2}^{Q} z_n | n \rangle \big],
\end{split}
\end{equation}

\noindent where $\Delta E = E_2 - E_1$ is the spectral gap. 
Observe that the linear combination $|\Phi_0 \rangle + c_1 |\Phi_1 \rangle \in K_{\hat{U}}(\Phi_0;N_T = 1)$ with a choice of $c_1 = - \exp{\left( i  E_2 t_1 \right)}$  (upon absorbing the global phase $\exp{(iE_1 t_1)}$) simply returns a scalar multiple of our target state $| 1 \rangle$.
Therefore VQPE converges exactly after one step for unstructured search, as long as we avoid specific timesteps that cause phase rotation by integer multiples of $2 \pi$. We note that this result does not change the analysis of unstructured search problem in regards to previous bounds established in the literature~\cite{Grover}. The creation of the state from which the flagged state can be sampled comes with a low probability of success, at least when employing a linear combination of unitaries (LCU) type of preparation, due to the low target amplitude typically limited to the initial state.~\cite{LCU}
Moreover, if the number of flagged states is unknown, relevant matrix elements have to be calculated on quantum computer.

\subsection{\label{sec:GroundState_Computation} Exponentially fast convergence of the ground state of a harmonic spectrum}
\begin{figure}[H]
\centering
\includegraphics[width=.4\textwidth]{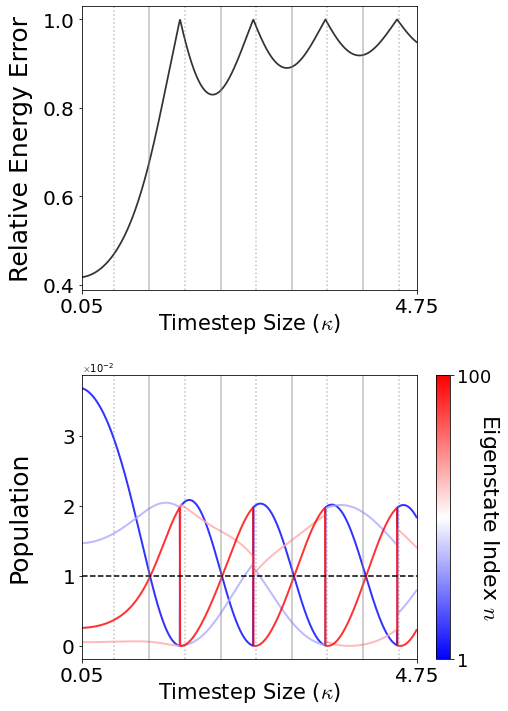}
\caption{Dependence of eigenstate population on the single timestep (linear spectrum $E_n = n \Delta E$). 
\textbf{Top}: The ground state energy error $\delta E_{1} = \langle \Psi_{\rm g} | \hat{H} | \Psi_{\rm g} \rangle - E_1$ is plotted over the timestep size $t_1\Delta E= \kappa 2\pi/Q$. Here $\delta E_1$ is plotted relative to the initial error $ \langle \Phi_0 |\hat{H}|\Phi_0\rangle - E_1$ and thus takes a value between $0$ (exact recovery of ground state) and $1$ (no improvement over the initial estimate). \textbf{Bottom}: The eigenstate population $p_n = |\langle n | \Psi_{\rm g} \rangle|^2$ from Eq.~\eqref{eq:pharmonic} is plotted over the timestep size $t_1\Delta E = \kappa 2\pi/Q$. Color within the lower panel distinguishes the eigenstates $|n\rangle$ and interpolates between blue ($n=1$) and red ($n=Q$). 
The black dashed line marks the initial population $p_n\equiv 1/Q$.}
\label{fig:Single_Timestep}
\end{figure}

In fact, we could regard the solution to the search problem as the ground states of many-body Hamiltonian operators found in condensed matter physics and quantum chemistry. In previous work we specifically examined the ground state wavefunctions of molecular Hamiltonians~\cite{Klymko2021_VQPE}. Instead, here we aim to understand the single step performance for some basic yet important model Hamiltonians. We first consider the Hamiltonian with a linear spectrum, 
\begin{eqnarray}
     \hat{H} = \sum_{n=1}^{Q} n \Delta E | n \rangle \langle n |,~\Delta E >0,
     \label{eq:Hamiltonian}
\end{eqnarray}
characteristic of a harmonic oscillator as a ubiquitous model in quantum mechanics. 
For normalized initial state $|\Phi_0 \rangle$, a single-step VQPE solves the linear equations introduced in Eq.~\eqref{eq:linsystem} where
\begin{eqnarray}
     &\mathbf{H} = \begin{bmatrix}
     \langle \Phi_0 | \hat{H} | \Phi_0 \rangle  
     & \langle \Phi_0 | \hat{H} \hat{U}(t_1) | \Phi_0 \rangle \\
     \langle \Phi_0 | \hat{U}(-t_1) \hat{H} | \Phi_0 \rangle
     & \langle \Phi_0 | \hat{H} | \Phi_0 \rangle \\
     \end{bmatrix}, \\
     &\mathbf{S} = \begin{bmatrix}
     1 & \langle \Phi_0 | \hat{U}(t_1) | \Phi_0 \rangle \\
     \langle \Phi_0 | \hat{U}(-t_1) | \Phi_0 \rangle & 1 \\
     \end{bmatrix},
\end{eqnarray}
give the $2 \times 2$ Hamiltonian and overlap matrix. For purpose of implementation, we choose $|\Phi_0\rangle$ to be the uniform superposition over eigenbasis $(z_n \equiv 1/\sqrt{Q})$ so it gets mapped to,
\begin{eqnarray}
     \begin{split}
         | \Phi_0 \rangle = | \Phi_{\rm U} \rangle &= \sum_{n=1}^{Q} \frac{1}{\sqrt{Q}} |n \rangle \\
         &\mapsto | \Phi_1 \rangle = \sum_{n=1}^{Q} \frac{\exp{\left( -i E_n t_1 \right)} }{\sqrt{Q}} |n \rangle,
     \end{split}
     \label{eq:Phi0_Phi1}
\end{eqnarray}
under time evolution. Let $| \Psi_{\rm g}(t_1) \rangle$ denote the ground state estimate, including a parametric dependence on the evolution time $t_1$. To solve the subspace-projected eigenvalue problem of Eq.~\eqref{eq:linsystem}, we identify a change-of-basis matrix $\mathbf{B}$ that orthogonalizes the overlap matrix ($\mathbf{B}^{\dagger} \mathbf{S} \mathbf{B} = \mathbf{I}$). This allows us to transform our original problem into the conjugated form of $\mathbf{B}^{\dagger} \mathbf{H} \mathbf{B} \vec{d} = E_{\rm g} \Vec{d}$. Once obtaining the solution in the new basis, we undo the basis change $\Vec{c} = \mathbf{B} \Vec{d}$, mounting to a weighted sum in the real-time basis $\ket{\Psi_{\rm g}} = \sum_{j} c_j \ket{\Phi_j}$. For a single-step evolution, it is straightforward to show that
\begin{equation}
     \hspace{-0.285cm} \mathbf{B} = \frac{1}{\sqrt{2}}  \begin{bmatrix}
        \frac{\mathbf{S}_{01} / \abs{\mathbf{S}_{01}}}{ \sqrt{1 + \abs{ \mathbf{S}_{01} } }} &  \frac{\mathbf{S}_{01} / \abs{\mathbf{S}_{01}} }{ \sqrt{1 - \abs{ \mathbf{S}_{01} } }} \\
        \\
        \frac{1}{\sqrt{1 + \abs{ \mathbf{S}_{01} } }} & \frac{- 1}{\sqrt{1 - \abs{ \mathbf{S}_{01} } }} \\
        \end{bmatrix}, ~~\Vec{d} = \begin{bmatrix}
         d_1 \\
         d_2 \\
        \end{bmatrix},
\end{equation}
with $d_1^{\ast} = - d_1$ and $d_2^{\ast} = d_2$, hence implying $\abs{c_0} = \abs{c_1}$. Assuming that $t_1$ satisfies a mild condition (discussed in Appendix \ref{sec:AppGroundState_Computation}), we can analytically derive the eigenstate population after the single timestep,
\begin{eqnarray}
     \begin{split}
        p_{n} &= \abs{\langle n | \Psi_{\rm g}(t_1) \rangle}^2 = \abs{ \frac{c_0 + c_1\exp{(-i E_n t_1)}}{\sqrt{Q}} }^2, \\
        &= \frac{ 2 \Re \left[c_0^{\ast} c_1\exp{(-i E_n t_1)} \right] + \abs{c_0}^2 + \abs{c_1}^2 }{Q},\\
        &= \frac{\sin{\left[ \rchi(t_1) +  E_n t_1 \right]} + 1}{\mathcal{Z}(t_1)},
     \end{split}
     \label{eq:pharmonic}
\end{eqnarray}

\noindent where $\rchi$ and $\mathcal{Z}$ represent, respectively, some phase offset and normalization constant determined by the matrix elements $\mathbf{H}_{ij}$ and $\mathbf{S}_{ij}$ (specified in Appendix \ref{sec:AppGroundState_Computation}). 
The sinusoidal dependence in Eq.~\eqref{eq:pharmonic} is explicitly displayed in Fig.~\ref{fig:Single_Timestep} with an optimal timestep $\Delta E t_1 \in (0, \pi/Q)$ observed.

\begin{figure}[t!]
\centering
\includegraphics[width=.35\textwidth]{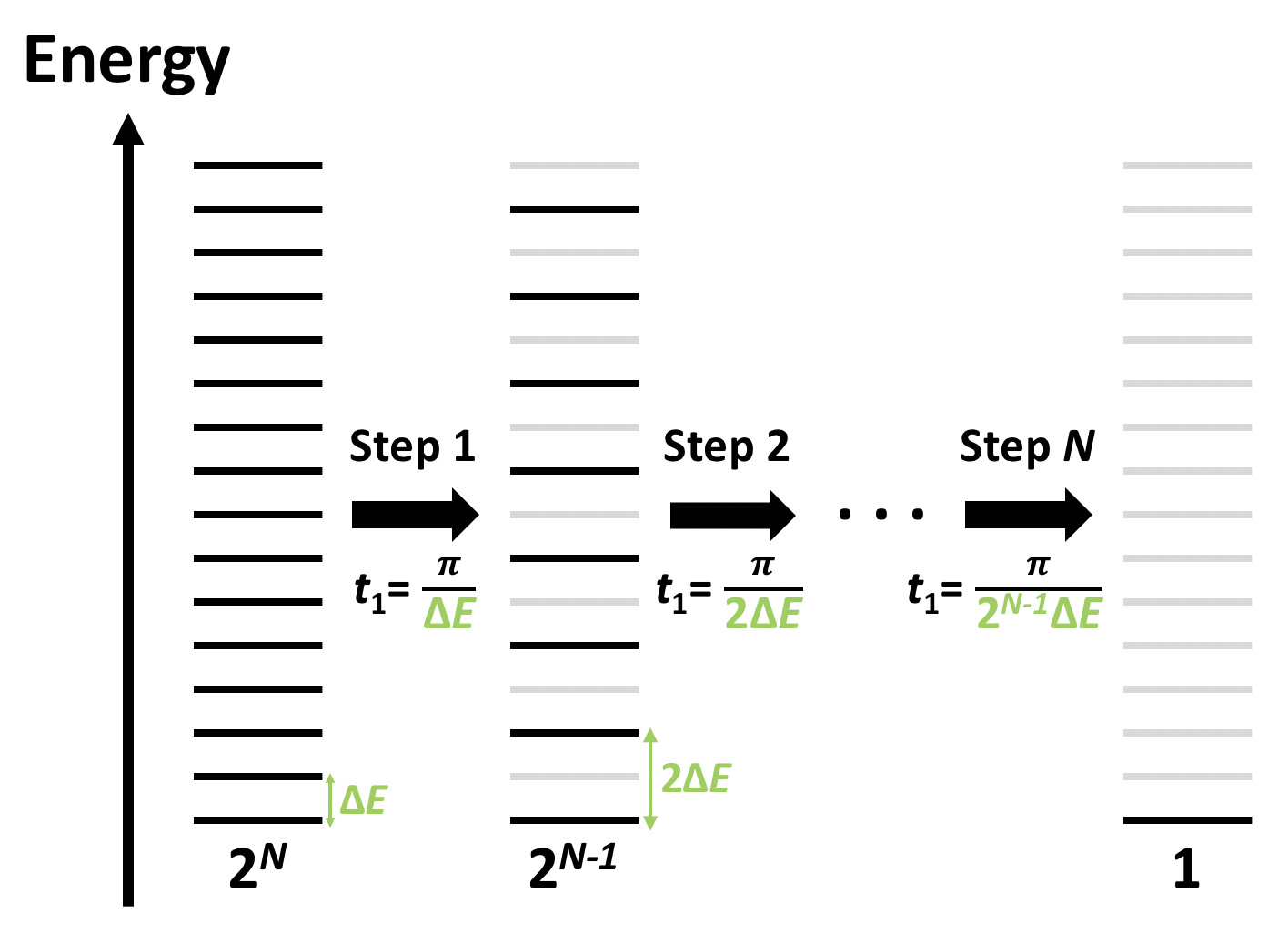}
\caption{Exponential convergence to the ground state from a sequence of single step real-time evolutions. The eigenstate population is shown throughout the evolutions, starting from the initial state of uniform superposition $\ket{\Phi_{\rm U}}$. At each step, constructive and destructive phase interference are colored by black and grey respectively, where the exponentially decreasing number of support eigenstates is labelled at the bottom.}
\label{fig:HO_exp_convg}
\end{figure}

For the special case $\Delta E t_1 = \pi$ and $Q \in 2\mathbb{Z}^{+}$, Eq.~\eqref{eq:pharmonic} simplifies such that the extracted ground state becomes,
\begin{eqnarray}
    | \Psi_{\rm g} (\pi/\Delta E) \rangle = \sum_{1 \leq n \leq Q}^{n~{\rm odd}} \sqrt{\frac{2}{Q}} | n \rangle,
    \label{eq:HO_exp_convg}
\end{eqnarray}
with half of the population amplitude eliminated and the other half doubled due to constructive and destructive interference. In fact, such interference readily underlies our earliest example of a single spin subject to the Pauli-$Z$ Hamiltonian in Sec.~\ref{sec:Single step}, where we trivially recognize that $\ket{\downarrow} = \ket{\Psi_{\rm g}(\pi/2)}$ for $\Delta E = 2$ and $Q=2$.
The simple result of Eq.~\eqref{eq:HO_exp_convg} implies that given a linear spectrum, exact recovery of the ground state in a Hilbert space of dimension $2^N$ only takes a sequence of $N$ single steps (once we recalibrate the excitation energy $\Delta E \mapsto 2 \Delta E$ after each step). This recovery procedure is schematically illustrated in Fig.~\ref{fig:HO_exp_convg}, exhibiting a particular yet clear manifestation of the exponential convergence from real-time evolution.

As a final note on the convergence properties from a single step evolution, we also examine in Appendix~\ref{sec:Single-Step Gap} a minimally modified Hamiltonian based on Eq.~\eqref{eq:Hamiltonian}, featuring a variable spectral gap, denoted as $\epsilon_{12}$, between the ground and first excited state. Predictably, we find that a change in spectral gap induces a population transfer among the eigenstates, where the lower energy population can be enhanced by a larger gap. In the extreme case of infinitely large gap, we effectively recover unstructured search for which the VQPE solution becomes exact after a single timestep.

\subsection{\label{sec:Random_SingleStep} Continuum modeling of spectrum}

\begin{figure}[H]
\centering
\includegraphics[width=.375\textwidth]{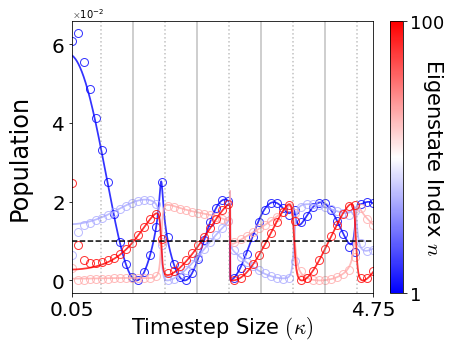}
\caption{Eigenstate population from given spectral density. Eigenstate population $p_n$ is plotted over timestep size $t_1\Delta E= \kappa 2\pi/Q$ for a gapped linear spectrum $E_n = n \Delta E + (1-\delta_{1,n})\epsilon_{12}$ with $\epsilon_{12}= 20 \Delta E$ (solid line) and for a spectrum with flat spectral density $\omega(E)$ (hollow circles). Color distinguishes the eigenstates $|n\rangle$ and interpolates between blue ($n=1$) and red ($n=Q$). The black dashed line marks the initial population $p_n\equiv 1/Q$.}
\label{fig:Single_Rand}
\end{figure}

In the large $Q$ limit, we may treat the spectrum as some continuum with a prescribed density of states (DOS) that reflects the probability of observing a certain energy level.
We remark that the single step expression of Eq.~\eqref{eq:pharmonic} remains valid for arbitrary spectrum $\{E_{n}\}_{n=1}^{Q}$, and a spectrum dilation $E_n \mapsto c E_n$ preserves the eigenstate population up to a stretch of time $t_1 \mapsto t_1/c$. %%KK question here
Consequently, we assume that the spectral range, $E_Q - E_1$, is bounded from above by some fixed finite constant $C > 0$. 
For sufficiently large $Q$, we simply approximate the Hamiltonian and overlap matrix elements via a properly normalized spectral density of states $\omega(E)$, \textit{i.e.},
\begin{eqnarray}
     \int_{E_1}^{E_1 + C} \omega(E) dE = Q,
\end{eqnarray}
so that,
\begin{eqnarray}
     \sum_{n=1}^{Q} f(E_n;t_1) \approx \int_{E_1}^{E_1 + C} f(E;t_1) \omega(E)dE ,
     \label{eq:Rand_Approx}
\end{eqnarray}
for relevant functions $f$ of energy. Specifically,
\begin{eqnarray}
     \begin{split}
        \mathbf{H}_{01} &\approx \int_{E_1}^{E_1 + C} E\exp{(-i E t_1)}\omega(E)dE, \\ 
        \mathbf{S}_{01} &\approx \int_{E_1}^{E_1 + C} \exp{(-i E t_1)} \omega(E)dE,
     \end{split}
\end{eqnarray}
where the $f$-integrals evaluate to the characteristic function $\hat{\omega}(t_1)$ of the DOS and its first derivative.
Eq.~\eqref{eq:Rand_Approx} establishes the real-time subspace on the mean level via the approximation $(\mathbf{H}, \mathbf{S}) \mapsto (\mathbb{E} \mathbf{H}, \mathbb{E} \mathbf{S})$, where the mean $\mathbb{E}$ is taken over the joint spectral distribution $\omega^{(Q)}(E_1, \cdots, E_Q) $ defined through
\begin{equation}
     \omega(E) = \int \prod_{n} d E_n~ \omega^{(Q)} (\{E_n\}) \sum_{n} \delta (E - E_n).
\end{equation}
The sum-integral relation from Eq.~\eqref{eq:Rand_Approx} holds precisely if $\omega(E)$ is the empirical DOS of a discrete spectrum. 
As an illustrative example, we compare a linear spectrum and a spectrum with uniform DOS in Fig.~\ref{fig:Single_Rand}. The population profiles show reasonable agreement as expected, and the difference that emerges at short evolution time will vanish in the large $Q$ limit.

\subsection{Locating spectral suppression}

The spectral location of the characteristic suppression seen in Fig.~\ref{fig:singlestep} can be calculated by extremizing Eq.~\eqref{eq:pharmonic}, 
\begin{eqnarray}
   \begin{split}
       \frac{\partial p(E)}{\partial E}  &= 0, \\
       \frac{\partial^2 p(E)}{\partial^2{E}} &> 0,
   \end{split}
\end{eqnarray}
where the population profile $p(E; t_1)$ and its derivatives are understood from our continuum modeling  (Sec.~\ref{sec:Random_SingleStep}) in the large $Q$ limit.
Let us scale our spectrum to a range of $[0,1]$ so that the resulting suppression occurs around $E_{x} = (1-x)E_1 + x E_Q$ for some $x$ specifying the center of the suppressed region. For a linear spectrum $E_n = n \Delta E$ and suitably short evolution,  
\begin{eqnarray}
     x = -\frac{\lim_{t_1 \rightarrow 0} \rchi'(t_1) + \Delta E}{(Q-1) \Delta E} \approx 0.8.
\end{eqnarray}
independent of the spectral spacing $\Delta E$ and evolution time $t_1$, which is consistent with our observation.
\bigskip

\section{\label{sec:Multi-step convergence}Beyond Single Step}
\subsection{\label{sec:Multi-step}Multi-step convergence}

\begin{figure}[tbh!]
\includegraphics[width=.475\textwidth]{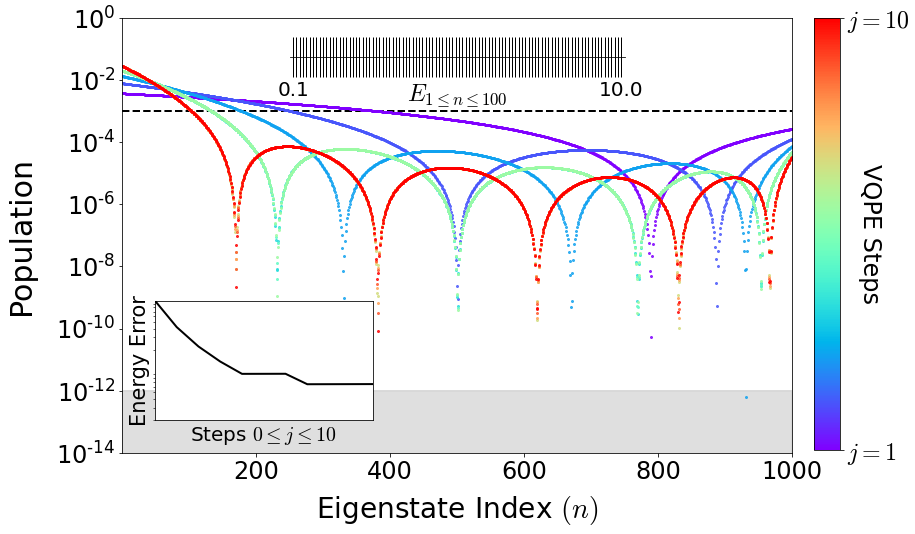}
\includegraphics[width=.475\textwidth]{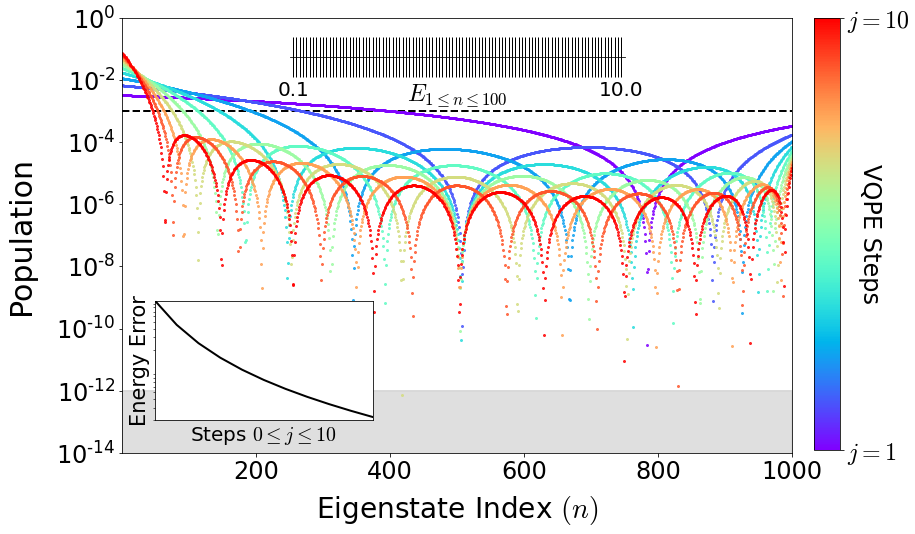}
\includegraphics[width=.475\textwidth]{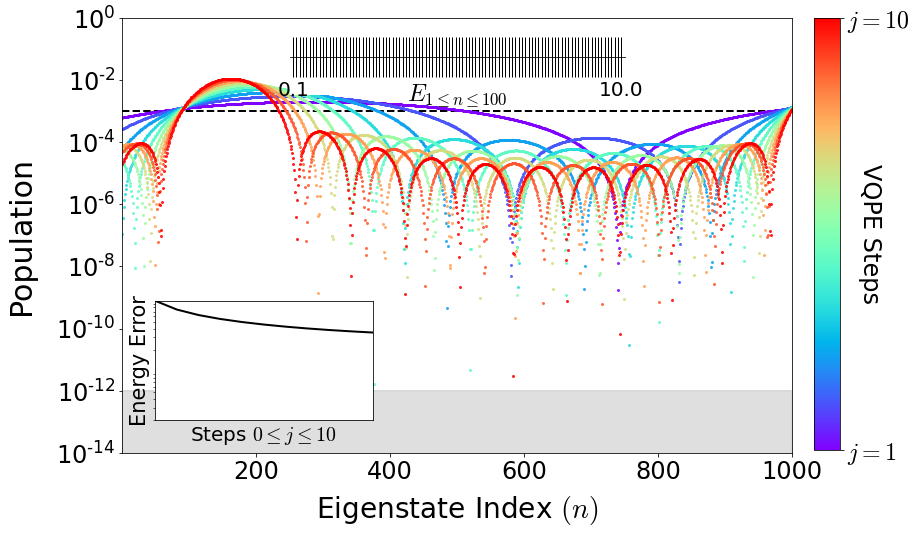}
\caption{Structured population suppression over the excited states after multiple timesteps (linear spectrum $E_n = n \Delta E$ and linear time grid $t_j = j t_1$). Population profiles $p_{n,j}$ are plotted as a function of the eigenstate index $1 \leq n \leq Q = 1000$ for three different values of timestep size $t_1\Delta E = \kappa 2\pi/Q$. \textbf{Top}: Small timestep $\kappa = 0.05$. \textbf{Middle:} Moderate timestep $\kappa = 0.4$. \textbf{Bottom}: Large timestep $\kappa = 1.1$.  Profile color in all panels indicates the number of timesteps taken and interpolates between purple ($j=1$) and red ($j=10$). The insets display the log-scale energy error $\delta E_{1,j} = \langle \Psi_{\rm g}(j) | \hat{H} | \Psi_{\rm g}(j) \rangle - E_1$ as the number of timesteps increases.}
\label{fig:Mutiple_Timestep_Compare}
\end{figure}

VQPE leads to an exponential suppression of the excited state population as we take more timesteps. 
In particular, a multi-step evolution facilitates delocalized spectral decays, where the number of decay centers over the spectrum grows in proportion to the number of timesteps. 
Such structured suppression of the eigenstate population $p_{n, j} = |\langle n | \Psi_{\rm g}(t_1, \cdots, t_j) \rangle|^2$ can be visualized in Fig.~\ref{fig:Mutiple_Timestep_Compare}. 

For multi-step VQPE, fast convergence relies on a suitable time grid $\vec{t}$. We want the real-time states $\{ |\Phi_{j} \rangle \}_{j=0}^{N_T}$ to be sufficiently independent in the sense that the singular values $s_j \in [0, N_T + 1]$ of the overlap $\mathbf{S}$ stay bounded, for example, below by a threshold value $s_{\rm SV}$. In practice, we simply solve Eq.~\eqref{eq:linsystem} on a truncated subspace for which $s_{j} \geq s_{\rm SV}$ to avoid numerical instabilities~\cite{Higham1998} and filter out noise. The insets within Fig~\ref{fig:Mutiple_Timestep_Compare} show the convergence measured by the ground state energy error. 
Observe that a small timestep introduces linear dependency in the real-time states and slows down the convergence, which is also manifested in the population profile.
On the other hand, a large timestep deteriorates the evolution by introducing degeneracies in the phase interference pattern and hence undesirably suppressing the low energy population. This happens when
\begin{eqnarray}
     (E_n - E_1) t \geq 2\pi,
\end{eqnarray}
where large $t$ values result in phase wrappings around the origin in the complex plane. 
Clearly the condition above imposes periodicity in eigenstate population, where degeneracies arise in the phases accumulated by eigenstates that are nonadjacent in the spectrum. Therefore leveraging the two notions of independence, we may cleverly choose the timesteps so that VQPE converges the fastest. 
Such optimal time choice for recovering the ground state differs from that investigated in the previous work for recovering the full spectrum.

\begin{figure}[tbh!]
\centering
\includegraphics[width=.375\textwidth]{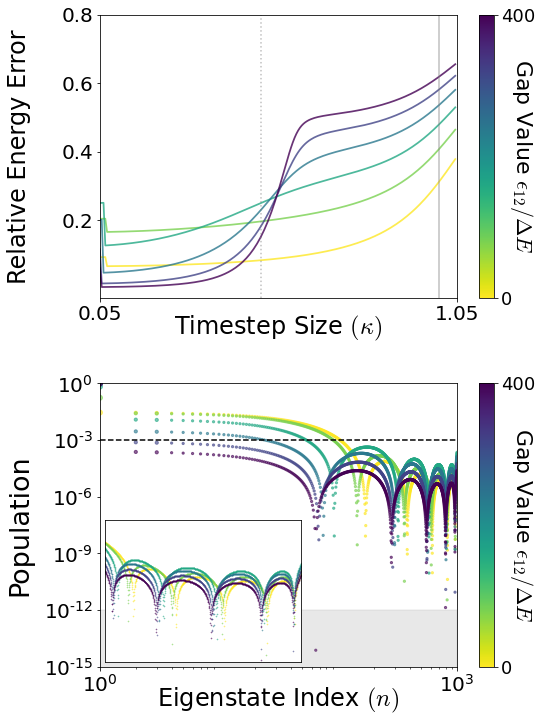}
\caption{Structured population suppression over excited states after multiple timesteps (linear spectrum with spectral gap $E_n = n\Delta E + (1-\delta_{1,n})\epsilon_{12}$ and time grid $t_j = j t_1$). \textbf{Top}: Ground state energy error $\delta E_{1,j} = \langle \Psi_{\rm g}(j)| \hat{H} | \Psi_{\rm g}(j) \rangle - E_1$  with $j=5$ is plotted as a function of the timestep size $t_1 \Delta E = \kappa 2\pi/ Q$. The relative energy error is normalized by the initial error $ \langle \Phi_0 |\hat{H}|\Phi_0\rangle - E_1$ and determines the optimal timestep size. \textbf{Bottom}: The resulting eigenstate population from the optimal timestep is plotted over the eigenstate index $1\leq n\leq Q=1000$. Curve color distinguishes the gap value and interpolates between yellow ($\epsilon_{12}=0$) and dark green ($\epsilon_{12}=400\Delta E$). The inset zooms over the population decays in the observed profile.}
\label{fig:Mutiple_Timestep_Gap}
\end{figure}

The ground state convergence also admits a native dependence on the spectral gap $\epsilon_{12}$. 
In the single step limit, the presence of a spectral gap changes the location of the population suppression. 
For a gapped linear spectrum and suitably short evolution, $E_{x} = - \lim\nolimits_{t_1 \rightarrow 0} \rchi'(t_1|\epsilon_{12})$ determines the suppressed energy in the spectrum from Eq.~\eqref{eq:pgap}. 
Notice that $E_{x}$ is naturally associated with an eigenstate $| n_1 \rangle$ for which
\begin{eqnarray}
     n_1 \approx -\frac{\lim\nolimits_{t_1 \rightarrow 0} \rchi'(t_1|\epsilon_{12}) + \epsilon_{12}}{\Delta E},
\end{eqnarray}
monotonically decreases with the gap value. Therefore one expects a red shift of the decay center $n_1(\epsilon_{12})$ relative to $n_1(0) \approx x Q$ if $\epsilon_{12} > 0$ and a blue shift otherwise. In either situation, the shift originates from the additional phase separation $\exp{(-i\epsilon_{12}t_1)}$ between the ground and first excited state. Borrowing our intuition from the single step limit, we expect a larger spectral gap to red shift and broaden the decay regions in a multi-step simulation, as is shown in Fig~\ref{fig:Mutiple_Timestep_Gap}. Accompanied with the red shift is a faster convergence, since a larger gap better separates the excited state phases from the ground state phase. Effectively, the higher energy phases are squeezed together so that they undergo more thorough phase cancellations. As $\epsilon_{12} \rightarrow \infty$, a single step suffices to recover the ground state as already discussed in Secs~\ref{sec:Grover} and \ref{sec:GroundState_Computation}.

\subsection{Proof of multi-step convergence}
Even for real-time evolution employing a simple linear time grid, the error of our spectral approximation can be bounded based on an extension of the Kaniel–Paige–Saad formalism~\cite{parlett1980_eigvalbound,Paige1971_thesis, Saad1980_cor,lin2022qsd}. In particular, we establish an error bound through the following theorems.
\\

\noindent \textbf{Theorem 1.1.} Let $E_{\Tilde{1}}(j)$ label the approximate lowest eigenvalue within the subspace $K_{\hat{U}}(\Phi_0; j)$, and $\delta E_{1}(j) = E_{\Tilde{1}}(j) - E_{1}$ the energy error. Then for $j \geq 1$, there exists time grid spacing $\Delta t$ such that,
\begin{eqnarray}
    0 \leq \delta E_{1}(j) \leq \frac{ (E_{Q} - E_{1}) \Tilde{\epsilon}_{1,2}^{-2j} \sin^2{\Xi}}{ \cos^2{\Xi} },
    \label{ineq:loeigval_bound}
\end{eqnarray}
where $\cos^2{\Xi} = |\langle \Phi_0 | 1 \rangle|^2$ measures the squared overlap between the initial state and the true ground state while $\Tilde{\epsilon}_{1,2} = 1 + 3(E_2 - E_1)\Delta t/2\pi$ $\in [1,2]$ characterizes the normalized spectral gap.

\begin{figure}[H]
\centering
\includegraphics[width=.425\textwidth]{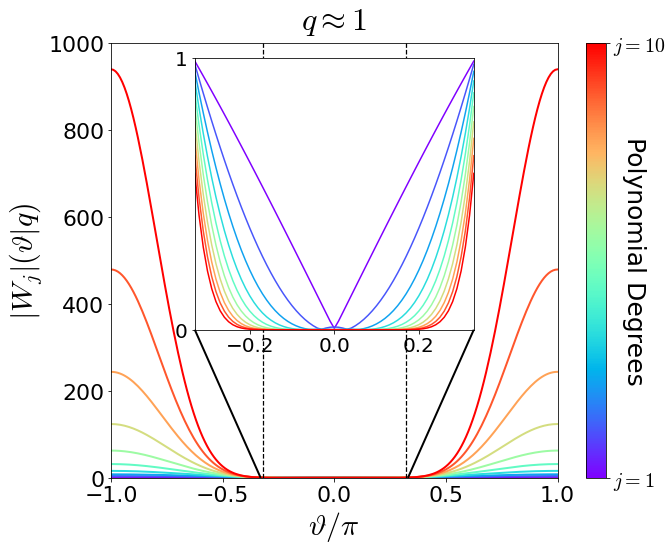}
\caption{Rogers-Szeg\H{o} polynomials of varying degrees. The modulus $|W_{j}(z| q)|$ of the polynomials is plotted as a function of the angular phase variable $\vartheta$ with $q \approx 1$ fixed (recall that $z = -q^{-1/2}\exp{(-i\vartheta)}$). Curve color indicates the degree of a polynomial and interpolates linearly between purple ($j=1$) and red ($j=10$). Inset illustrates the bounded behavior of $W_j$ over the angular window $[-\pi/3, \pi/3]$ marked by the vertical dashed lines in the main plot.}
\label{fig:RS_Polys}
\end{figure}

\noindent[\textit{proof}.] Let us define,
\begin{eqnarray}
       r(v) =  \frac{\langle v | \hat{H} | v \rangle}{\langle v | v \rangle},
       \label{eq:E_expect}
\end{eqnarray}
which returns the expected energy of state $|v\rangle$.
We focus on the rightmost inequality since the left simply restates,
\begin{eqnarray}
    E_1 &=& \min\nolimits_{ |v \rangle \neq 0 \in \mathcal{H}} r(v), \\
    &\leq& \min\nolimits_{ |v \rangle \neq 0 \in K_{\hat{U}}(\Phi_0; j) } r(v) = E_{\Tilde{1}}(j).
\end{eqnarray}
Notice that up to a spectral flip $\hat{H} \mapsto - \hat{H}$, it suffices to prove the equivalent statement on $\delta E_{Q}$ for which a more natural argument is entailed. By definition,
\begin{eqnarray}
    \begin{split}
        E_{\Tilde{Q}}(j) &= \max_{ |v \rangle \in K_{\hat{U}}(\Phi_0; j)} \frac{\langle v | \hat{H} | v \rangle }{\langle v | v \rangle}, \\
        &= \max_{p \in \mathcal{P}_{j} } \frac{\langle \Phi_0 | p(\hat{U})^{\dagger} \hat{H} p(\hat{U})  | \Phi_0 \rangle}{\langle \Phi_0  p(\hat{U})^{\dagger} p(\hat{U}) | \Phi_0  \rangle},
    \end{split}
    \label{eq:hieigval}
\end{eqnarray}
for which $\mathcal{P}_{j}$ denotes the set of degree $j$ polynomials over $\mathbb{C}$ and $\hat{U} \equiv \hat{U}(\Delta t)$. 
Although yet to be identified, we know that there exists a unique set of coefficients $\{ z_n \}_{n=1}^{Q}$ of $|\Phi_0\rangle$ with respect to the true eigenbasis $\{ |n\rangle \}_{n=1}^{Q}$ such that the expression above can be rewritten as,
\begin{eqnarray}
    \begin{split}
        |\Phi_0 \rangle =& \sum_{n=1}^{Q} z_n |n \rangle \\
        &\implies E_{\Tilde{Q}} = \max_{p \in \mathcal{P}_{j}} \frac{ \displaystyle \sum_{n=1}^{Q} E_{n} \abs{z_n p(\lambda_n)}^2 }{ \displaystyle \sum_{n=1}^{Q} \abs{z_n p(\lambda_n)}^2},
    \end{split}
    \label{eq:hieigval_2}
\end{eqnarray}
where $\lambda_{n} = \exp{(- i E_n \Delta t)}$ and we have exploited the unitarity $\hat{U}^{\dagger} = \hat{U}^{-1}$ so that $p(\hat{U})^{\dagger} |n\rangle = p(\lambda_n)^{\ast} |n\rangle$ with $\ast$ denoting the complex conjugation. Relaxing the numerator in Eq.~\eqref{eq:hieigval_2}, we have
\begin{eqnarray}
    \hspace{-1.5 cm} E_{\Tilde{Q}} &\geq& \max_{p \in \mathcal{P}_j} \frac{ \displaystyle E_Q \abs{z_Q p(\lambda_Q)}^2 + E_1 \sum_{n=1}^{Q-1} \abs{z_n p(\lambda_n)}^2 }{ \displaystyle \sum_{n=1}^{Q} \abs{z_n p(\lambda_n)}^2 }, \\
    &=& E_{Q} - (E_Q - E_1) \min_{p \in \mathcal{P}_j} \frac{ \displaystyle \sum_{n=1}^{Q-1} \abs{z_n p(\lambda_n)}^2 }{ \displaystyle \sum_{n=1}^{Q} \abs{z_n p(\lambda_n)}^2 }.
     \label{eq:hieigval_3}
\end{eqnarray}
In the original Kaniel–Paige–Saad formalism, an advantageous choice of $p \in \mathcal{P}_j$ that realizes a tight bound is the real-valued Chebyshev polynomials,
\begin{eqnarray}
     T_{j}(x) = \begin{cases}
     1         & j = 0 \\ 
     x         & j = 1 \\
     2x T_{j-1}(x) - T_{j-2}(x) & j \geq 2
     \end{cases},
\end{eqnarray}
where the minimal supremum norm property of $T_{j}$, 
\begin{eqnarray}
     \frac{1}{2^{j-1}} \norm{T_j(x)}_{\infty} = \inf_{p \in \mathcal{P}_j : p - x^j \in \mathcal{P}_{j-1} } \norm{p_{}(x)}_{\infty},
\end{eqnarray}
over the interval $[-1, 1] \subset \mathbb{R}$ helps establish the suitable bound on the fraction in Eq.~\eqref{eq:hieigval_3}. Note that here $\lambda_n = \exp{(-i\vartheta_n)}$ for $\vartheta_n = E_n \Delta t \in [0, 2\pi)$ so we seek a family of polynomials defined over the unit circle $\mathbb{S}^1 = \{ z: |z| = 1 \} \subset \mathbb{C}$ to bound the fraction. Let $q \in (0,1]$ and now we will consider the handy choice of complex-valued Rogers-Szeg\H{o} polynomials~\cite{RSpoly,orthogonalpoly}, 
\begin{eqnarray}
      W_j(z|q) = \begin{cases}
      1         & j = 0 \\ 
      z + 1     & j = 1 \\
      (1+z) W_{j-1}(z|q) \\
      \hspace{.6 cm} - (1-q^{j-1}) z W_{j-2}(z|q) & j \geq 2
      \end{cases},
      \label{eq:R-S}
\end{eqnarray}
over the circle $\mathbb{S}^{1,q} = \{z: |z| = q^{-1/2} \}$. For simplicity, we rewrite $z = -q^{-1/2}\exp{(-i\vartheta)}$ where $\vartheta \in [-\pi, \pi)$ denotes an angular phase. A prefactor of $-1$ is included to periodically translate the polynomials so that $W_j(\vartheta|q)$ adapts the symmetry $W_j(-\vartheta)=W_j(\vartheta)^{\ast}$ (we also omit a conditional dependence of $W_j$ on $q$ for notational clarity). Such family of polynomials shares the key properties that $(i)$ $|W_j(\vartheta)|$ remains bounded below unity over some proper angular window $\mathcal{W} = [-\Omega, \Omega]$ $\subset [-\pi, \pi)$ and $(ii)$ $|W_j(\vartheta)|$ grows rapidly outside $\mathcal{W}$. For explicit illustrations, Rogers-Szeg\H{o} polynomials of the first few orders are shown in Fig.~\ref{fig:RS_Polys}.

\medskip

Note that the constant $q$ controls the width of our truncated angular window $\mathcal{W}$. In the limit $q \rightarrow 1$, one can verify that these polynomials converge to, 
\begin{eqnarray}
      W_j(\vartheta) \rightarrow \sum_{k=0}^{j} \binom{j}{k} \exp{\left[ -ik(\vartheta+\pi) \right]},
\end{eqnarray}
which simply gives the sum of evenly spaced points on $\mathbb{S}^1$ weighted by the binomial coefficients. As a consequence, $\sup_{\vartheta} |W_j(\vartheta)| \approx2^j$ for $q \approx 1$. To bound the fraction from Eq.~\eqref{eq:hieigval_3} tightly, we want a suitable linear transformation $\mathcal{L}$ acting on the eigenphases $\{\lambda_{n}\}_{n=1}^{Q}$ such that $\mathcal{L}$ nudges $\vartheta_{n \leq Q-1}$ all inside the truncated window $\mathcal{W}$ while keeping $\vartheta_Q$ outside. Without loss of generality, we may assume $\vartheta_{Q-1} - \vartheta_1 \leq 2\Omega$ and $ 2\Omega \leq \vartheta_{Q} - \vartheta_1 \leq \pi + \Omega$  by choosing suitable time grid $\Delta t$, \textit{e.g.},
\begin{eqnarray}
      \begin{split}
          \hspace{-0.5cm} \Delta t = \sup_{\tau} \Big\{ \tau \in \mathbb{R}^{+}: &\vartheta_{Q}(\tau) - \vartheta_{Q-1}(\tau) \leq \Omega^{\rm c}, \\
          &\vartheta_{Q-1}(\tau) - \vartheta_{1}(\tau) \leq 2\Omega \Big\},
      \end{split}
\end{eqnarray}
with $\Omega^{\rm c} = \pi - \Omega$. Hence a natural $\mathcal{L}$ is the phase multiplicative transformation,
\begin{eqnarray}
     \mathcal{L}: \vartheta \mapsto \vartheta + \Omega - \vartheta_{Q-1},
     % \mathcal{L}: z \mapsto \exp{\left[i(\vartheta_{Q-1} - \Omega) \right]} z
\end{eqnarray}
which circularly shifts $\{\lambda_{n}\}_{n=1}^{Q}$ so $|\mathcal{L}(\vartheta_{1})| \leq \mathcal{L}(\vartheta_{Q-1}) = \Omega \leq \mathcal{L}(\vartheta_{Q})$ as desired. With our pick of $\mathcal{L}$, we can establish a variational upper bound by substituting the trial polynomials $p = W_j \circ \mathcal{L}$ into Eq.~\eqref{eq:hieigval_3},
\begin{eqnarray}
       \hspace{-0.85 cm} \inf_{\Delta t} E_{Q} - E_{\Tilde{Q}} &\leq& (E_Q - E_1) \frac{\displaystyle \sum_{n=1}^{Q-1} \abs{z_n W_j(\Omega) }^2 }{ \abs{z_{Q} W_j (\mathcal{L}(\lambda_{Q}))}^2},\\
       &=& (E_Q - E_1) \frac{\sin^2{\Xi}}{\abs{W_j (\mathcal{L}(\lambda_{Q}))}^2 \cos^2{\Xi}},
       \label{eq:bound1}
\end{eqnarray}
\smallskip

\noindent where in arriving at Eq.~\eqref{eq:bound1} we have utilized property $(i)$ of $W_j$ and defined an overlap angle $\Xi$ by $\cos^2{\Xi} = |z_{Q}|^2 = |\langle \Phi_0 | Q \rangle|^2$ that specifies the projection of our initial state onto the top eigenstate. For the limiting case $q = 1$, it is rather straightforward to show that $\Omega = \pi/3$ and,
\begin{eqnarray}
     \begin{split}
          \abs{W_j \left( \mathcal{L}(\lambda_Q) \right)}^{1/j} &= \sqrt{2 - 2\cos{\left( \vartheta_{Q} - \vartheta_{Q-1} + \Omega \right)}}, \\
          &\geq  1 + \Gamma \epsilon_{Q-1,Q},
     \end{split}
     \label{eq:Linbound}
\end{eqnarray}
where $\epsilon_{Q-1,Q} = (\vartheta_{Q} - \vartheta_{Q-1})/\Omega^{\rm c}$ denotes the normalized top spectral gap and $\Gamma$ is a constant for which Ineq.~\eqref{eq:Linbound} holds for $\epsilon_{Q-1,Q} \in [0, 1]$. For example, $\Gamma = 1$ is justified by concavity of the LHS of the inequality with respect to the spectral gap $\vartheta_{Q}-\vartheta_{Q-1}$. Hence we can further bound Eq.~\eqref{eq:bound1} using Ineq.~\eqref{eq:Linbound},
\begin{eqnarray}
      \inf_{\Delta t} E_{Q} - E_{\Tilde{Q}} \leq \frac{ (E_Q - E_1) \Tilde{\epsilon}_{Q-1,Q}^{-2j} \sin^2{\Xi} }{ \cos^2{\Xi}}, 
      \label{eq:bound2}
\end{eqnarray}
for $\Tilde{\epsilon}_{Q-1,Q} = 1+\Gamma \epsilon_{Q-1,Q} \geq 1$. After flipping $\hat{H} \mapsto -\hat{H}$, we have proved the statement in the theorem as claimed. Notice that our result is analogous to the classical Krylov result except that $\Tilde{\epsilon}_{Q-1,Q} = 1 + 2 \epsilon_{Q-1,Q} + 2(\epsilon_{Q-1,Q}^2 + \epsilon_{Q-1,Q})^{1/2}$ with $\epsilon_{Q-1,Q} = (E_{Q} - E_{Q-1})/(E_{Q}-E_{1})$ was used in the original convergence theory~\cite{parlett1980_eigvalbound,Paige1971_thesis, Saad1980_cor}. $\square$

\bigskip

\noindent \textbf{Corollary 1.2.} Let $E_{\Tilde{n}}(j)$ label the approximate $n$th lowest eigenvalue and $\delta E_{n}(j) = E_{\Tilde{n}}(j) - E_{n}$ the energy error. Then for $j \geq n \geq 1$, there exists time grid $\Delta t$ such that,
\begin{eqnarray}
    0 \leq \delta E_{n}(j) \leq \frac{ (E_{Q} - E_{n}) Y_{n,j} \Tilde{\epsilon}_{n, n+1}^{-2(j-n+1)} \sin^2{\Xi_n}}{ \cos^2{\Xi_n} },
    \label{ineq:loeigval_bound2}
\end{eqnarray}
where $Y_{n,j}$ is a prefactor containing the $(n-1)$ lowest approximations,
\begin{eqnarray}
      Y_{n,j} = \begin{cases}
      1 & n =1 \\
      \displaystyle \max_{ \ell > n } \prod_{m=1}^{n-1} \abs{ \frac{\lambda_{\ell} - \exp{\left( -i E_{\Tilde{m}} \Delta t \right)}}{ \lambda_n - \exp{\left(  -i E_{\Tilde{m}} \Delta t \right)} } } & n \geq 2 \\ 
      \end{cases},
\end{eqnarray}
while $\cos^2{\Xi_n} = |\langle \Phi_0 | n \rangle|^2$ and $\Tilde{\epsilon}_{n,n+1} = 1 + 3(E_{n+1} - E_n) \Delta t/2\pi$ denote the relevant squared overlap and interior spectral gap respectively. Recall that we have defined the phase factors $\lambda_{\ell} = \exp{\left( - i E_{\ell} \Delta t \right)}$ asscoiated with the true eigenvalues in Thm 1.1.

\noindent[\textit{proof}.] Again we present the argument for $\delta E_{Q-n+1}$ due to the identification $E_{n} \leftrightarrow E_{Q-n+1}$ through a spectral flip. For simplicity, let $\star n = Q-n+1$. First observe that by the min-max characterization of operator eigenvalues~\cite{horn} as embodied in Eq.~\eqref{eq:E_expect},
\begin{eqnarray}
     \hspace{-0.75 cm} E_{\Tilde{\star n}}(j) - E_{\star n} &=& \max_{R \subseteq K_{\hat{U}} } \min_{|v \rangle \in R} r(v|\hat{H} - E_{\star n} \hat{I}), \\
     &\leq& \max_{R \subseteq \mathcal{H}} \min_{|v\rangle \in R} r(v|\hat{H} - E_{\star n} \hat{I}) = 0,
\end{eqnarray}
where $\hat{I}$ denotes the identity operator and $R$ labels an $n$-dimensional subspace. Thus it suffices to establish RHS of Ineq.~\eqref{ineq:loeigval_bound2}. By construction $\delta E_{\star n}  \geq \max_{\mathcal{M}} r(v|\hat{H} - E_{\star n} \hat{I}) \implies - \delta E_{\star n} \leq \min_{\mathcal{M}} r(v|E_{\star n} \hat{I} - \hat{H})$ given $\mathcal{M} = {\rm span}\left\{ |\Tilde{m} \rangle \right\}_{m=Q-j}^{\star n} \subset K_{\hat{U}}(\Phi_0;j)$, so we have
\begin{eqnarray}
      \hspace{-0.5cm} \abs{\delta E_{\star n}} &\leq& \min_{|v \rangle = p(\hat{U}) |\Phi_0 \rangle} \frac{\displaystyle \sum_{\ell=1}^{Q} (E_{\star n} - E_{\ell}) \abs{z_{\ell} p(\lambda_{\ell})}^2 }{\displaystyle \sum_{\ell=1}^{Q} \abs{z_{\ell} p(\lambda_{\ell})}^2 }, \\
      &\leq& \min_{p(\hat{U})|\Phi_0 \rangle} \frac{\displaystyle \sum_{\ell=1}^{\star n - 1} (E_{\star n} - E_{\ell}) \abs{z_{\ell} p(\lambda_{\ell})}^2}{\displaystyle \abs{z_{\star n} p(\lambda_{\star n})}^2 },
      \label{ineq:eigerror_interior}
\end{eqnarray}
where the minimum is taken over the subset of polynomials $p \in \mathcal{P}_j$ satisfying $\langle \Tilde{m} | p(\hat{U}) | \Phi_0 \rangle = 0$ for $\star n + 1 \leq m \leq Q$ (we have reserved the same notations as in Thm.1.1). Here we extend Saad's main idea and consider reducible complex polynomials of the form,
\begin{eqnarray}
      p(z) &=& q(z) \prod_{m = \star n +1}^{Q} \frac{ \displaystyle z -  \exp{\left(-i E_{\Tilde{m}} \Delta t \right)} }{ \displaystyle \lambda_{\star n} - \exp{\left( - i E_{\Tilde{m}} \Delta t \right)} }, \\
      &=& q(z) p_{\downarrow}(z),
\end{eqnarray}
with $p$ factorizable into two polynomials $q \in \mathcal{P}_{j-n+1}$ and $p_{\downarrow} \in \mathcal{P}_{n-1}$. By design, the complex exponentials $\{\exp{(-i E_{\Tilde{m}} \Delta t)} \}_{m=\star n +1}^{Q}$ are zeros of $p$ so $p(\hat{U}) |\Phi_0 \rangle \in \mathcal{M}$ is guaranteed with $(n-1)$ orthogonality conditions above fulfilled. On the other hand, $p_{\downarrow}(\lambda_{\star n}) = 1$ implies,
\begin{eqnarray}
      \abs{\delta E_{\star n}} \leq \min_{q} (E_{\star n} - E_1) \frac{\displaystyle \sum_{\ell=1}^{\star n - 1} \abs{p_{\downarrow}(\lambda_{\ell}) z_{\ell} q(\lambda_{\ell})}^2}{\displaystyle \abs{z_{\star n} q(\lambda_{\star n})}^2 },
\end{eqnarray}
and we may simply relax the numerator by recognizing,
\begin{eqnarray}
      \abs{p_{\downarrow}(\lambda_{\ell})} \leq \sup_{z \in \mathcal{A}(\star n ;\Delta t)} \abs{p_{\downarrow}(z)},
\end{eqnarray}
where $\mathcal{A}(\star n ; \Delta t) \subset \mathbb{S}^{1}$ gives a circular arc with arc angle $[-\vartheta_{\star n - 1}, -\vartheta_{1}]$. For example, we expect the supremum to occur at $\lambda_1 = \exp{(-i \vartheta_1)}$ when the time grid $\Delta t$ satisfies $\vartheta_{Q} - \vartheta_1 \leq \pi$. It is clear that the rest of our proof follows from direct application of Thm 1.1 to the spectral sector $\{E_{\ell}\}_{\ell=1}^{\star n}$. $\square$

\section{\label{sec:Implementation analysis} Alternative Implementation analysis}
\subsection{\label{sec:Vanilla iterative} Vanilla and iterative time evolution}
VQPE evolving a fixed reference $|\Phi_0 \rangle$ for $N_T$ timesteps solves a linear system in one shot, which requires $\mathcal{O}(N_{T}^{\beta})$ complexity with an exponent $ \beta \in [2,3]$. Now consider an evolution for which we dynamically update the reference after each timestep. Specifically, we update based on our current best guess $| \Psi_{\rm g}(j) \rangle$ on an $(N_{I}+1)$-dimensional subspace defined iteratively by,
\begin{eqnarray}
     {\rm span}\left\{\hat{U}(\Delta t_{j,k} )|\Psi_{\rm g}(j-1) \rangle: 0 \leq k \leq N_I \right\},
     \label{eq:iterative_subspace}
\end{eqnarray}
where $t_{j-1} = t_{j,0} < t_{j,1} < \cdots < t_{j, N_I} = t_j $ denotes a partition of $[t_{j-1}, t_j]$ with $\Delta t_{j,k} = t_{j,k} - t_{j,k-1}$ ($\Delta t_{j,0} \equiv 0$ as our convention). 
The trivial case $N_T = 1$ corresponds to a vanilla VQPE. For the simplest nontrivial case $N_I = 1$, the reference state after $j$ steps takes the form,
\begin{equation}
     |\Psi_{\rm g}(j) \rangle \propto |\Psi_{\rm g}(j-1) \rangle + c_{j} \hat{U}(\Delta t_j) |\Psi_{\rm g}(j-1) \rangle,
\end{equation}
starting with $| \Psi_{\rm g}(0) \rangle = | \Phi_0 \rangle$, our input initial state. Such iteration requires $\mathcal{O}(N_T N_I^\beta)$ complexity and thus offers a speedup when $N_I \ll N_T$. In this section, we focus on the case $N_I = 1$, again with our initial state being a uniform superposition. 

\begin{figure}[H]
\centering
\includegraphics[width=.4\textwidth]{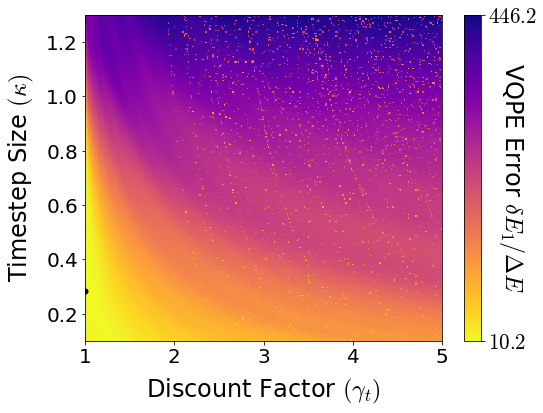} \includegraphics[width=.4\textwidth]{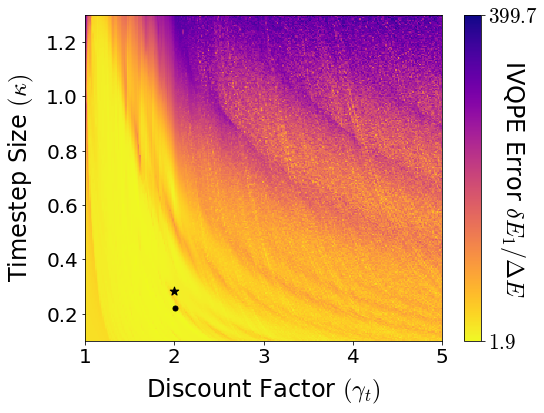}
\caption{Convergence of ground state energy for evolution of $N_T = 10$ adaptive timesteps (linear spectrum $E_n = n \Delta E$). Ground state energy error $\delta E_{1,N_T} = \langle \Psi_{\rm g}(N_T) | \hat{H} | \Psi_{\rm g}(N_T) \rangle - E_1$ is plotted in units of $\Delta E$ as a function of two parameters $\kappa = t_1 Q \Delta E / 2\pi$ and $\gamma_t \geq 1$. \textbf{Top}: Nondimensional energy error from vanilla evolution (VQPE). Filled circle in black highlights the  optimal parameters. \textbf{Bottom}: Nondimensional energy error from iterative evolution (IVQPE). Filled circle in black highlights the optimal parameters and filled star marks the near optimal parameters $(t_1, \gamma_t) = (t_1^{\star}, 2)$.}
\label{fig:IVQPE_adapted}
\end{figure}

First observe that a vanilla time evolution always outperforms an iterative one if we employ a linear time grid, $\vec{t} = (t_1, 2t_1, \cdots, N_T t_1)$, based on the update rule from Eq.~\eqref{eq:iterative_subspace}. As a minimal example, $|\Psi_{\rm g}(j=2) \rangle$ takes the free form $c_0 |\Phi_0 \rangle + c_1 | \Phi_1 \rangle + c_2 | \Phi_2 \rangle$ under a vanilla evolution and the constrained form $d_0(c_0|\Phi_0 \rangle + c_1 |\Phi_1\rangle) + d_1(c_0|\Phi_1 \rangle + c_1 |\Phi_2 \rangle)$ under an iterative evolution. Instead, we consider the adaptive time grid,
\begin{eqnarray}
     \vec{t} = (t_1, \gamma_t t_1 + t_1, \cdots, \sum_{j=1}^{N_T} \gamma_{t}^{j-1} t_1),
     \label{eq:timegrid_ada}
\end{eqnarray}
where $\gamma_t$ defines an additional free parameter discounting %%KK
any time interval $[t_{j-1}, t_{j}]$ with respect to its precursor $[t_{j-2}, t_{j-1}]$. Now $|\Psi_{\rm g}(j=2) \rangle$ from the example above takes the form $d_0(c_0|\Phi_0 \rangle + c_1 |\Phi_1\rangle) + d_1(c_0|\Phi_2 \rangle + c_1 |\Phi_3 \rangle)$ after an iterative evolution with $\gamma_t = 2$, thus including a new state $|\Phi_3\rangle$ that can potentially facilitate the convergence. 
For subsequent comparisons, let $t_1^{\star}$ denote the size of a timestep from the optimal linear grid. 
We then look at convergence of the iterative VQPE (IVQPE) with an adaptive time grid versus the vanilla VQPE with an optimal linear time grid $t_j = j t_1^{\star}$. When $\gamma_t \ll 1$, both vanilla and iterative evolution degrade with more timesteps due to linear dependency issues. 
When $\gamma_t > 1$, we expect the iterative evolution to gain reasonable convergence at $(t_1, \gamma_t) = (t_1^{\star}, 2)$ which we term near optimal parameters, since we iteratively evolve onto a larger unexplored subspace of size $2^j$ that stays maximally independent from the explored one. 
The near optimality over our restricted two-dimensional parameter space is explicitly displayed in Fig.~\ref{fig:IVQPE_adapted}. 
% We note that $\gamma_t = 2$ remains a hardware-efficient choice on quantum device since it only demands measurements linear in the number of timesteps. 

\begin{figure}[H]
\centering
\includegraphics[width=.475\textwidth]{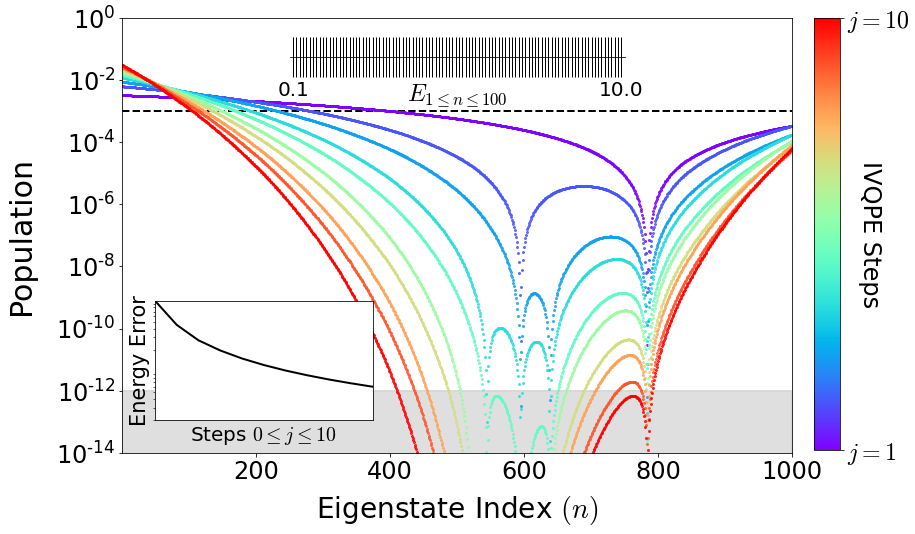}
\includegraphics[width=.475\textwidth]{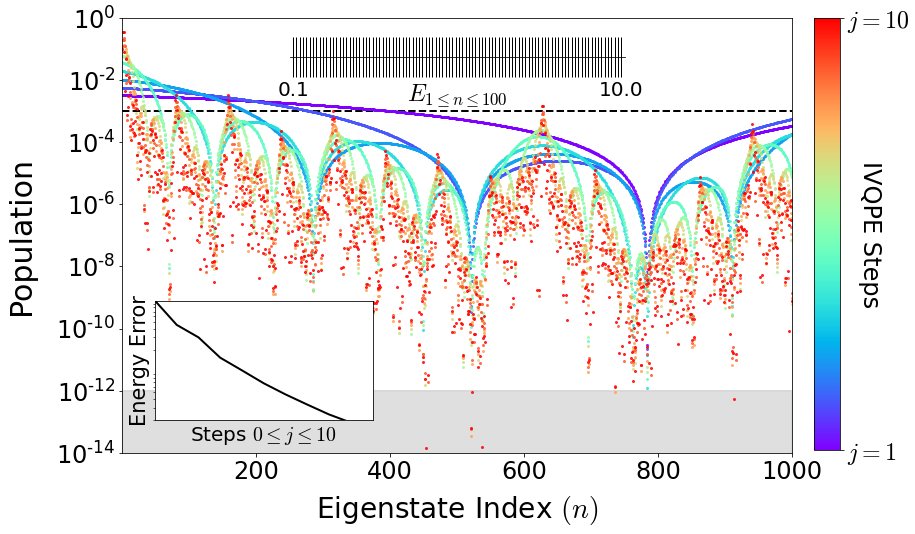}
\caption{Population suppression over excited states from a multi-step iterative evolution (linear spectrum $E_n = n \Delta E$). Population profiles are plotted as a function of the eigenstate index $1 \leq n \leq Q = 1000$ for different time grid parametrizations with $t_1\Delta E = \kappa 2\pi/Q$. \textbf{Top}: A linear time grid $\Delta t_j = t_1$ with $\kappa =  0.4$. \textbf{Bottom}: An adaptive time grid $\Delta t_j = {\gamma_t}^{j-1} t_1$ with $(\kappa,\gamma_t) = (0.4, 2)$. In both panels, profile color indicates the number of iterative timesteps taken and interpolates between purple ($j=1$) and red ($j=10$). The inset displays the ground state energy error $\delta E_{1,j} = \langle \Psi_{\rm g}(j) | \hat{H} | \Psi_{\rm g}(j) \rangle - E_1$ on a log scale as a function of the steps taken.}
\label{fig:Mutiple_Timestep_Compare_VI}
\end{figure}

%%KK what does near optimal mean?
The population profiles of ground states extracted from IVQPE with linear and adaptive time grids are displayed in Fig.~\ref{fig:Mutiple_Timestep_Compare_VI}, where we observe very different population suppression depending on the time grid that guides the phase rotations. 
For $0 \ll \gamma_t < 1$, we expect the performances of both vanilla and iterative evolution to progressively degrade as $\gamma_t$ decreases, where the vanilla evolution will experience a sharper slowdown in its convergence rate due to difficulties in simultaneously resolving all the time evolved states for desirable phase cancellation. 
However, the use of $\gamma_t < 1$ turns out to be particularly beneficial for special scenarios. 
For example, recall that in Sec.~\ref{sec:GroundState_Computation}, we have deduced an exact and exponentially fast ground state recovery from an iterative evolution with adaptive timesteps $\Delta t_{j} = 2^{1-j} \pi / \Delta E$ and thus $\gamma_t = 1/2$. 
In that case, specific restrictions lie in the spectral density (linear spectrum) and Hilbert space size ($\log_{2}Q \in \mathbb{Z}^{+}$).

%\textcolor{blue}{
The implementation of IVQPE as a quantum circuit via a sequence of intermediate state preparation is discussed in Appendix \ref{sec:IVQPE_Preparation}. We remark that an accurate sampling of the Hamiltonian and overlap matrix elements relies on the faithful preparation of the intermediate states. 
Regardless of the time parametrization, we only need to measure the off-diagonal matrix elements (the diagonal elements are determined before each IVQPE step). For the simplest case $N_{I} = 1$, the total number of measurements is $2N_T$, \textit{i.e.}, still linear in the number of timesteps taken.
%}

\subsection{\label{sec:Stochasticity effect}Effect of stochasticity}

Within the context of ground state computation from real-time evolution, sources of stochasticity may include dynamical noises due to dissipative system-bath interactions and statistical uncertainties due to measurements on hardware, both of which evolve with the number of timesteps taken. 
By simulating perturbations on our target spectrum, we can examine susceptibility of the multi-step convergence to induced spectral disorder. 

\begin{figure}[htb!]
\centering
\includegraphics[width=.475\textwidth]{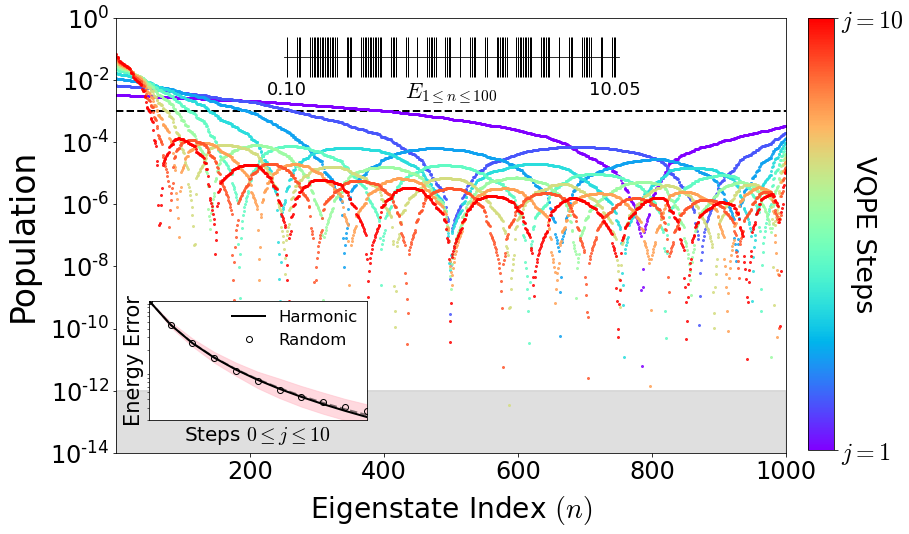}
\includegraphics[width=.475\textwidth]{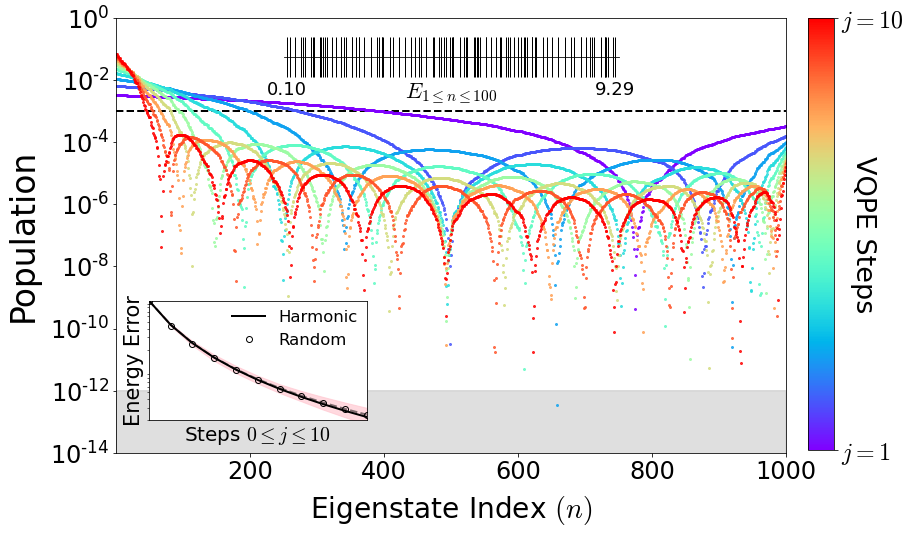}
\includegraphics[width=.475\textwidth]{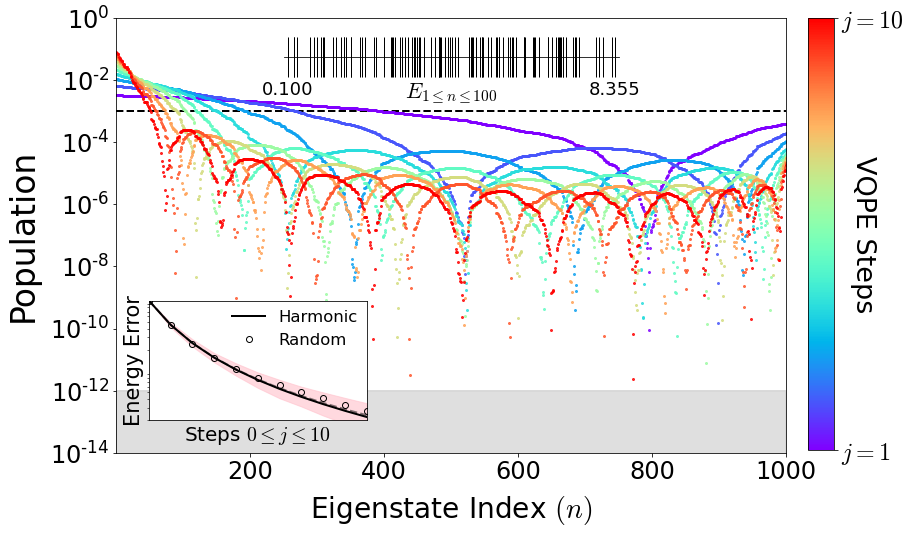}
\caption{Ground state convergence after multiple timesteps (random spectrum and linear time grid $t_j = j t_1$ with $\kappa = t_1 Q  \mathbb{E}\Delta E / 2\pi = 0.4$). Population profiles are plotted for three distributions of energy level spacing. \textbf{Top}: Bernoulli spacing. \textbf{Middle}: Uniform spacing. \textbf{Bottom}: Exponential spacing. 
The profile color in all panels indicates number of timesteps taken and interpolates between purple ($j=1$) and red ($j=10$).
The inset displays the log-scale ground state energy error $\delta E_{1,j}$ where the solid curve benchmarks the convergence for the linear spectrum while hollow circles mark the convergence for a randomly realized spectrum. 
The pink shade shows the convergence envelope for $10^3$ random realizations. }
\label{fig:Mutiple_Timestep_Random}
\end{figure}

Let us absorb such disorder into the spectral DOS $\omega(E)$ introduced in Sec.~\ref{sec:Random_SingleStep}. 
Without stochasticity, $\omega(E) = \sum_{n=1}^{Q} \delta(E- E_n)$ is a collection of sharp peaks in the energy domain. 
These peaks will broaden in the presence of probabilistic perturbations so that $\omega(E) = \sum_{n=1}^{Q} g_n(E)$, where the broadening is dictated by the distributions, $g_{n}$, from which the energy levels are drawn. %\textcolor{blue}{
For concreteness, a phenomenological instance of the spectral broadening is derived in Appendix \ref{sec:Noise_Model} using perturbation theory.%} 
Here we consider a random spectrum with fixed ground state energy $E_1$ and $i.i.d.$ level spacing,
\begin{eqnarray}
      E_{n+1} - E_{n} = \Delta_n \sim p_{\Delta}(\Delta_n),
     \label{eq:level_spacing_prob}
\end{eqnarray}
where $p_{\Delta}(\Delta_n)$ gives the spacing statistics. 
In this specific case, the distributions $g_n$ are given by,
\begin{eqnarray}
     \begin{split}
         g_n(E) = \int \prod_{m=1}^{n-1} d \Delta_m ~\delta(E - \sum_{m<n} &\Delta_m - E_1) \\
         &\prod_{m<n} p_{\Delta}(\Delta_m),
     \end{split}
\end{eqnarray}
with all the spacing variables $\Delta_m$ integrated out through convolutions of their statistical weights. 
From here on, we will use the single random variable $\Delta E \sim p_{\Delta}(\Delta E)$ to denote the level spacing in the presence of stochasticity. Upon averaging over the spectral disorder associated with $g_n$, we get an effectively linear spectrum, $E_n^{\rm eff} = E_1 + (n-1) \mathbb{E} \Delta E$. 
Note by the generalized Jensen's inequality,
\begin{eqnarray}
      \begin{split}
          \abs{\mathbb{E}^n \exp{(-i E \Delta t_j)} -\exp{(-i E_n^{\rm eff} \Delta t_j)}}& \\
          &\hspace{-2.25cm} \leq \frac{\eta (n-1) (\Delta t_j \mathbb{E}\Delta E)^2}{\sqrt{2}},
      \end{split}
\end{eqnarray}
where $\mathbb{E}^n$ denotes an expectation against the level distribution $g_n$ and $\eta > 0$ defines a nondimensional level spacing variance such that ${\rm var} \Delta E = \eta (\mathbb{E} \Delta E)^2$. 
Thereby for suitably short time evolution $\vec{t}$ (as remarked in Sec.~\ref{sec:Multi-step}) and spacing disorder $p_{\Delta}$ with controlled variance, a standard continuity argument suggests that the approximate eigenvalues of a target operator subject to stochastic perturbations above will, on average, closely resemble those of a deterministic operator having a harmonic spectrum. 
As a result, these perturbations can be significantly tempered through ensemble average over the spectral disorder, especially for unperturbed target operators with a relatively flat DOS. To illustrate this robustness of VQPE with respect to spectral disorder from Eq.~\eqref{eq:level_spacing_prob}, we display in Fig.~\ref{fig:Mutiple_Timestep_Random} the ground state convergence for random spectra and their linear equivalent, given simple forms of $p_{\Delta}$ with $\eta = \mathcal{O}(1) \ll Q$.

\begin{figure}[tbh!]
\centering
\includegraphics[width=.475\textwidth]{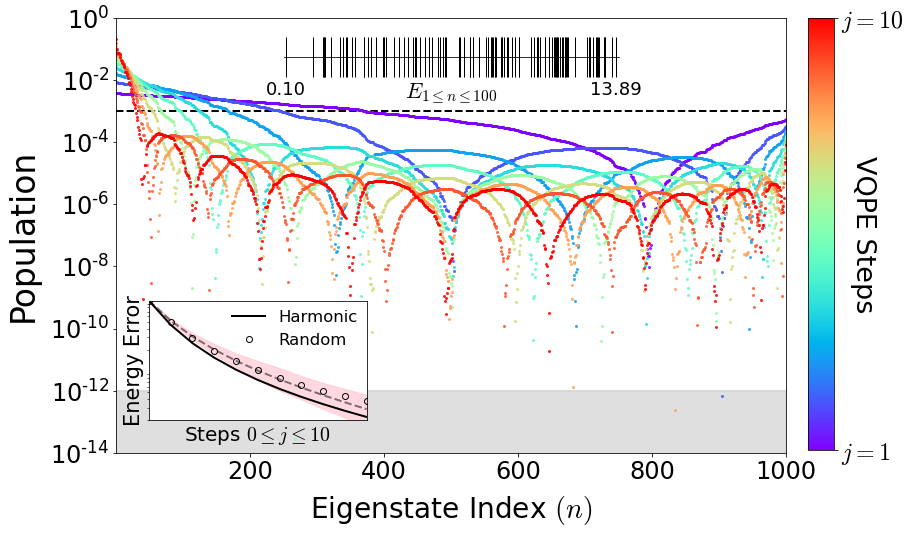}
\includegraphics[width=.475\textwidth]{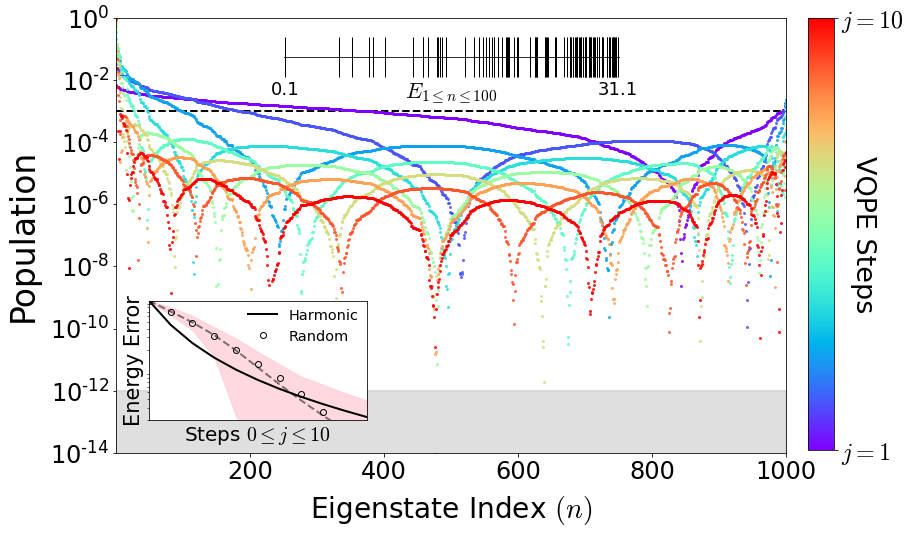}
\caption{Ground state convergence after multiple timesteps (random spectrum and linear time grid $t_j = j t_1$ with $\kappa = t_1 Q  \mathbb{E}\Delta E / 2\pi = 0.4$). Population profiles are plotted for two differently concentrated DOS $\omega(E)$. \textbf{Top}: Semicircular spectral density. \textbf{Bottom}: Gaussian spectral density. 
The profile color in both panels indicates the number of timesteps taken and interpolates between purple ($j=1$) and red ($j=10$). The inset displays the log-scale energy error where the solid curve benchmarks the convergence for a linear spectrum with a flat DOS while the hollow circles mark the convergence for a randomly realized spectrum. The pink shade shows the convergence envelope for $10^3$ random realizations and the dashed curve traces out the average convergence.}
\label{fig:Mutiple_Timestep_Random2}
\end{figure}

Agreement between the mean of the convergence envelope over spectral realizations and the convergence on the mean spectrum observed in Fig.~\ref{fig:Mutiple_Timestep_Random} tends to break down once the DOS $\omega(E)$ loses its dispersive character and builds up mass concentrations. 
For example, consider a random $Q \times Q$ Hermitian matrix with $i.i.d.$ $\mathcal{N}(0,1/Q)$ diagonal entries and $i.i.d.$ $\mathcal{N}(0, 1/2Q) + i \mathcal{N}(0, 1/2Q)$ upper-diagonal entries, \textit{i.e.},
\begin{eqnarray}
      \mathbf{H} \sim \frac{ \exp{\left[ - Q {\rm tr}(\mathbf{H}^2)/2 \right]} }{2^{Q/2} \pi^{Q^2/2}},
      \label{eq:GUE}
\end{eqnarray}
which generates the well-known Gaussian unitary ensemble (GUE) in random matrix theory. 
It is worth mentioning that the spectral disorder from Eq.~\eqref{eq:GUE} can be reproduced alternatively from the spacing statistics,
\begin{eqnarray}
      p_{\Delta}(\Delta E) = \frac{32(\Delta E)^2 \exp{\left[-4(\Delta E)^2/\pi D^2 \right]}}{\pi^2 D^3},
\end{eqnarray}
where $D = \mathbb{E}\Delta E$ and $p_{\Delta}$ in this case vanishes quadratically for small $\Delta E$, exhibiting the phenomenon of level repulsion. From either prescription, one may prove that the DOS follows a semicircle law $\omega(E) = Q\sqrt{4-E^2}/2\pi$ with mass concentrated around $E=0$. Such concentration can be contrasted with the dispersion $\omega(E) \propto 1/D$ when $p_{\Delta}(\Delta E) \propto \exp{(-\Delta E/D)}$. Fig.~\ref{fig:Mutiple_Timestep_Random2} demonstrates the convergence behavior of random spectra sampled within GUE  upon proper scaling and shifting, where we note a drift of the convergence envelope away from our convergence benchmark on the averaged spectrum $E_{n}^{\rm eff}$. 

To further investigate the dependence of the convergence envelope on the DOS concentration, we also include in Fig.~\ref{fig:Mutiple_Timestep_Random2} the behavior of random spectra sampled according to the Gaussian density $\omega(E) \propto \exp{(-E^2/2\sigma^2)}$ where $\sigma$ tunes the mass concentration. For a chosen mean spacing $\mathbb{E} \Delta E$, stochasticity with Gaussian DOS shows a faster ground state convergence compared to that with semicircular DOS as displayed in Fig.~\ref{fig:Mutiple_Timestep_Random2}. Thus the shape of $\omega(E)$ takes a decisive part in regulating the convergence of VQPE. In general, $\omega(E)$ is uniquely determined by its characteristic function $\hat{\omega}(t)$. But a suitably short time evolution allows us to extrapolate $\hat{\omega}(t)$ only in terms of the derivatives $\hat{\omega}^{(k)}(0)$, which are nothing but the cumulants of our DOS. Consequently, we comment that the different convergence behaviors due to different disorder may be exploited as a spectral fingerprint from which useful local information about the eigenvalue density can be revealed.

\subsection{\label{sec:Initial vector} Choice of initial vector}
Throughout the previous sections, we have asserted a simplifying assumption that we initialize with a uniform superposition state. 
In practice, this assumption seems tailored for certain tasks such as the combinatorial searches (creation of equally weighted bitstrings from Hadamard gates) but becomes less effective to implement for other tasks such as the electronic structure predictions in quantum chemistry. Specifically, in contrast to massively degenerate spectra whose eigenstates could be easily resolved within few timesteps as seen earlier in Sec.~\ref{sec:Grover}, molecular Hamiltonians generally exhibit a full-rank structure accompanied by, at most, moderate spectral degeneracies. Consequently, achieving convergence for full-rank chemical problems requires exploring larger subspace, as is typical for the canonical Krylov methods. Given that phase cancellation is only relevant within the support of a real-time state, it is favorable to strategically prepare the initial state in order to taper the high Hamiltonian rank and minimize the runtime of the real-time evolution.

Here we consider the transition metal dimer ${\rm Cr}_2$ (def2-SVP basis set~\cite{weigend2005a}, 30 orbitals and 24 electrons), where we restrict the simulation to the widely studied 30-orbital active space, as a prototypical molecular system that exhibits strong electronic correlations. We then examine the role of initial state preparation in the VQPE ground state computation. Due to implementation feasibility, we truncate the Hilbert space and employ only a subset of all the Slater determinants in the active space.
The determinants are chosen using the adaptive sampling configuration interaction (ASCI) algorithm~\cite{Tubman2016,Tubman2018a,Tubman2020}. 
This is an iterative selected configuration interaction approach that explores the Hilbert space and identifies the most important determinants for a ground state approximation, thus providing highly accurate and moderately sized truncations. 
The data shown below for ${\rm Cr}_2$ includes 4000 determinants, selected by taking a one million determinant Hilbert space truncation with ASCI and picking the 4000 determinants with the largest coefficients in the one million.
Although the full Hilbert space for ${\rm Cr}_2$ is much larger than what we have studied here, we remark that in previous work~\cite{Klymko2021_VQPE} we performed a finite-size effect study of VQPE by comparing the dynamics of progressively larger truncations of Cr$_2$, from one thousand to one million determinants, demonstrating that vastly different truncation sizes result in the same convergence.

\begin{figure}[tbh!]
\centering
\includegraphics[width=.425\textwidth]{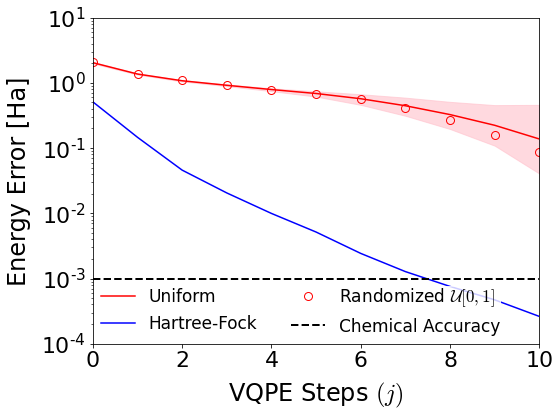}
\caption{${\rm Cr}_2$ ground state energy from a multi-step time evolution (adapted time grid $\Delta t_j = \gamma_t^{j-1} t_1 $ with $t_1$ and $\gamma_t$ optimized). Energy error is plotted over number of timesteps for different initial states where red and blue curves show the convergence for a uniform $| \Phi_{\rm U} \rangle$ and a Hartree-Fock $| \Phi_{\rm HF} \rangle$ initial state respectively. The red open circles mark the convergence for a randomized initial vector that approximates $|\Phi_{\rm U} \rangle$ and pink shade displays the convergence envelope for an ensemble of $10^3$ randomizations.The black dashed line indicates chemical accuracy.}
\label{fig:Cr2}
\end{figure}

We test two candidate initializations, uniform superposition $|\Phi_{\rm U} \rangle$ and Hartree-Fock  $| \Phi_{\rm HF} \rangle$, whose ground state convergences are shown in Fig.~\ref{fig:Cr2}. As the lower energy reference, the Hartree-Fock state also accelerates the rate of convergence and gives chemical accuracy on the order of $10$ timesteps. The observed fast convergence can be attributed to two primary features of $| \Phi_{\rm HF} \rangle$: $(i)$ its significant overlap with the low energy eigenstates, which persists across different system sizes and ensures an exponential decrease of the energy error, and $(ii)$ its relatively tight support in the Hilbert space, which effectively reduces the rank of the Hamiltonian and enlarges the spectral gap. Although this speedup necessitates a Hartree-Fock preparation beforehand, a handful of known techniques can be invoked to minimize the cost of such preprocessing so that $|\Phi_{\rm HF} \rangle$ remains an advantageous choice for eigenstate recoveries in molecular systems.

Moreover, we show in Fig.~\ref{fig:Cr2} the ground state convergence when the initial state is randomized as encountered in the standard Krylov subspace method, \textit{e.g.} the Lanczos algorithm. The randomization often involves a draw of $i.i.d.$ variables $\{ \phi_n \}_{n=1}^{Q}$ followed by a normalization $|\Phi_0 \rangle = (\phi_1, \cdots, \phi_Q) \mapsto |\Phi_0\rangle/\sqrt{\langle \Phi_0 | \Phi_0 \rangle}$. 
We take our drawing distribution to be uniform $\mathcal{U}[0,1]$ and plot the convergence envelope. 
By the central limit theorem, we expect the envelope to narrow and match the convergence behavior for $|\Phi_{\rm U} \rangle$ in the large $Q$ limit.
\bigskip

\section{Conclusion}
In this work, we study the class of subspace expansion algorithms utilizing a real-time evolved basis, providing detailed analysis for the underlying ideas in the original VQPE formalism from our previous work~\cite{Klymko2021_VQPE}. The main new results that we have presented here are the following: We have demonstrated a visualization of the convergence of the single-step and multi-step real-time algorithm. We have supplemented our visualization with a proof of the observed convergence, analogous to that constructed to justify Lanczos-type algorithms. Finally, we have introduced different algorithmic implementations for generating real-time states. This includes an iterative implementation that holds interesting convergence properties of its own. 
Given the significant recent interests in real-time algorithms on quantum devices~\cite{Stair2020,Parrish2019b,Klymko2021_VQPE,cortes2022,dong2022ground}, we believe that our work provides a timely and important step forward in understanding the properties of these algorithms as they are further developed for quantum computation. 
Our analysis, additionally, remains quite general and can also be used to advance these types of algorithms on classical architectures.
\bigskip

\section{Acknowledgments}
 %\sout{N.M.T., V.K.} 
  We are grateful for support from NASA Ames Research Center.   We acknowledge funding from the NASA ARMD Transformational Tools and Technology (TTT) Project.
  % When there are multiple funders, the DOE requires clear statements about what they funded versus what the other funders funded. This is an attempt to meet that requirement. 
  %YS acknowledges his material is based upon work supported by the U.S. Department of Energy, Office of Science, National Quantum Information Science Research Centers, Superconducting Quantum Materials and Systems Center (SQMS)under contract number DE-AC02-07CH11359.
  
  Calculations were performed as part of the XSEDE computational Project No. TG-MCA93S030 on  Bridges-2 at the Pittsburgh supercomputer center. 
  KK, DWY and WAdJ were supported by the Office of Science, Office of Advanced Scientific Computing Research Quantum Algorithms Team and Accelerated Research for Quantum Computing Programs of the U.S. Department of Energy under Contract No. DE-AC02-05CH11231.
  We are grateful to Daan Camps and Roel van Beeumen for discussions and careful readings of the manuscript.

\bigskip

\appendix
\section*{Appendix}
\section{\label{sec:Krylov_Basics} Krylov theory}
To elaborate on the theoretical footing that the Krylov method rests on, we introduce the Rayleigh quotient~\cite{Atkins2010}, 
\begin{eqnarray}
    r( v ) = \frac{\langle v | \hat{H} | v \rangle}{\langle v | v \rangle},
    \label{eq:Rayleigh quotient}
\end{eqnarray}
where $v$ represents a nonzero vector and we adopt Dirac's bra-ket notation in quantum mechanics to denote the inner product on $\mathcal{H}$. 
It is straightforward to check that the eigenvalue-eigenvector pairs $(E_{n}, |n\rangle)$ of the operator extremize the Rayleigh quotient in the sense that,
\begin{eqnarray}
    \begin{split}
        E_{n} &= \min\nolimits_{ |v \rangle \neq 0 \in \mathcal{H}_{n} } r(v), \\
        |n \rangle &= \argmin\nolimits_{ |v \rangle \in \mathcal{H}_{n}: \langle v | v \rangle = 1 } r(v),
    \end{split} 
    \label{eq:Rayleigh}
\end{eqnarray}
for which $\mathcal{H}_{n=1} = \mathcal{H}$ and $\mathcal{H}_{n\geq2} = {\rm span}\{| \ell \rangle: \ell \leq n-1 \}^{\perp}$ label the search space associated with the $n{\rm th}$ extreme eigenvalue.
As the dimensionality of the Hilbert space increases, the optimization task of exact operator diagonalization formulated in Eq.~\eqref{eq:Rayleigh} becomes numerically challenging despite active efforts to exploit existing Riemannian tools~\cite{OptMatrixManifold,SymplecticEigenpair,RiemannianImplicit} for tractable solutions. The Krylov method overcomes this numerical difficulty by restricting the optimization to the lower-dimensional Krylov subspace, 
\begin{eqnarray}
    K(\Phi_0; N_T) = {\rm span}\left\{ \hat{H}^{j} |\Phi_0 \rangle: j \leq N_{T} \right\},
\end{eqnarray}
for some initial vector $|\Phi_0 \rangle$ and number $N_{T}$ of repeated operator applications. The Krylov search space $K_{n} \subset K$, similar to $\mathcal{H}_{n}$, is defined recursively so that the resulting optimal eigenpairs $(E_{\Tilde{n}}, |\Tilde{n}\rangle)$ offer an approximation to the extreme ends of the spectrum. 
In the language of matrix algebra, minimization of the Rayleigh quotient restricted to the Krylov search space solves the equation,
\begin{eqnarray}
    \mathbf{H} \Vec{c}_{n} = E_{\Tilde{n}} \mathbf{S} \Vec{c}_{n}, 
\end{eqnarray}
where $\mathbf{H}$ and $\mathbf{S}$ denote the $\mathbb{C}^{{\rm dim}K \times {\rm dim}K}$ representation of the target and overlap operators in the Krylov basis,
\begin{eqnarray}
    \mathbf{H}_{ij} = \langle \Phi_i| \hat{H} | \Phi_j \rangle, ~ \mathbf{S}_{ij} = \langle \Phi_i | \Phi_j \rangle, 
\end{eqnarray}
and $\vec{c}_{n}$ denotes the $\mathbb{C}^{{\rm dim}K}$ coordinate of the vector $|\Tilde{n}\rangle$ in the same basis. 
In practice, the initial vector $|\Phi_0 \rangle$ can be chosen to ensure a full rank Krylov subspace of dimension $N_{T} + 1$.  This then makes the Krylov vectors linearly independent, and the Krylov basis can be orthonormalized by a modified Gram-Schmidt procedure such that $\mathbf{H}$ is tridiagonal and $\mathbf{S} \mapsto \mathbf{I}$ to avoid possible ill conditioning.

\section{\label{sec:Single-Step Gap} Influence of spectral gap on the single step convergence}
Let us consider the Hamiltonian with an additional spectral gap, 
\begin{eqnarray}
     \hat{H} = \Delta E | 1 \rangle \langle 1 | + \sum_{n=2}^{Q} \left( n \Delta E + \epsilon_{12} \right) | n \rangle \langle n |,
\end{eqnarray}
where $\epsilon_{12} \in (-\Delta E, \infty)$ denotes the signed excess excitation between the ground and first excited state. 
Unlike any spectrum shift which physically translates to a reset of the zero point energy and thus preserves the population (one may check that Eq.~\eqref{eq:pharmonic} remains invariant under energy shift $E_{n} \mapsto E_{n} + E_0$), a change in the spectral gap $\epsilon_{12}$  can induce a population transfer where the lower energy population gets enhanced by a larger gap value. Such nontrivial influence of the gap is manifested in the eigenstate population,
\begin{eqnarray}
     \hspace{-0.75cm} p_n = \frac{\sin{\left[\rchi(t_1| \epsilon_{12}) +  E_n t_1 \right]} + 1}{\mathcal{Z}(t_1 | \epsilon_{12})},
     \label{eq:pgap}
\end{eqnarray}
whose functional form is immediately accessible once we identify how the matrix elements transform under a gap change $E_n \mapsto E_n + (1 - \delta_{1,n}) \epsilon_{12}$.

\begin{figure}[H]
\centering
\includegraphics[width=.375\textwidth]{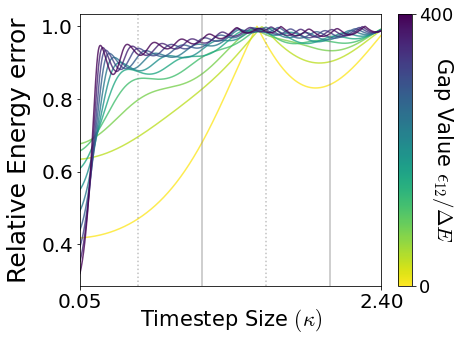}
\caption{Dependence of eigenstate population on the single timestep (linear spectrum with spectral gap $E_n = n\Delta E + (1-\delta_{1,n})\epsilon_{12}$). 
The ground state energy error $\delta E_{1} = \langle \Psi_{\rm g}| \hat{H} | \Psi_{\rm g} \rangle - E_1$ is plotted over the timestep size $t_1 \Delta E = \kappa 2\pi/ Q$.  Here $\delta E_1$ is plotted relative to the initial error $\langle \Phi_0 |\hat{H}|\Phi_0\rangle - E_1$ and takes a value between $0$ (exact recovery of ground state) and $1$ (no improvement over the initial estimate). Curve color indicates the value of the spectral gap and interpolates between yellow ($\epsilon_{12}=0$) and dark green ($\epsilon_{12}=400\Delta E$).}
\label{fig:Single_Timestep_Gap}
\end{figure}

Note that increasing $\epsilon_{12}$ in this case can enhance the population $p_1$, but at a likely cost of compromising energy accuracy in the single step limit. Such a trade-off is shown in Fig.~\ref{fig:Single_Timestep_Gap}. 
Even with the optimal timestep, the relative error in the approximate ground state energy (with respect to the equal superposition starting state) exhibits nonmonotonic dependence on the gap value. In the extreme case $\Delta E \rightarrow 0$ and $\epsilon_{12} \rightarrow \infty$, we recover the unstructured search for which the single step VQPE gives the exact result.

\section{\label{sec:AppGroundState_Computation} Exponentially fast convergence of the ground state of a harmonic spectrum}

\begin{figure}[tbh!]
\centering
\includegraphics[width=.375\textwidth]{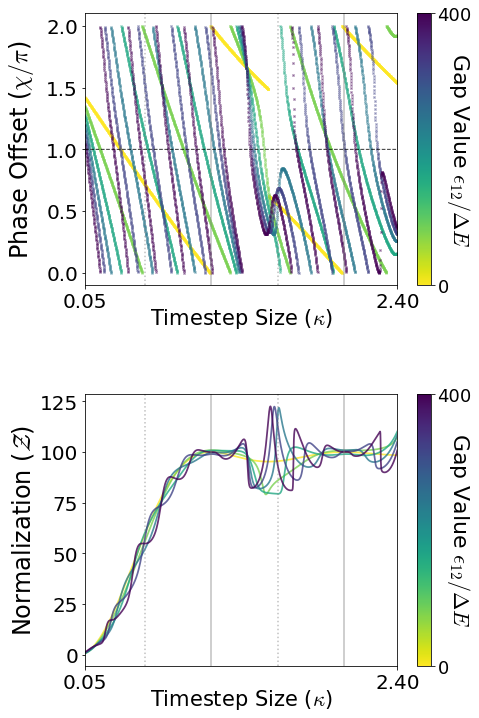}
\caption{Dependence of phase offset and amplitude normalization on the single timestep (linear spectrum with a gap $E_n = n\Delta E + (1-\delta_{1,n})\epsilon_{12}$). Curve color indicates the value of the spectral gap and interpolates between yellow ($\epsilon_{12}=0$) and dark green ($\epsilon_{12}=400\Delta E$). \textbf{Top}: Phase offset $\mathcal{\rchi}$ as a function of the timestep size $t_1 \Delta E = \kappa 2\pi/ Q$. \textbf{Bottom}: Normalization $\mathcal{Z}$ as a function of the timestep size $t_1 \Delta E = \kappa 2\pi/ Q$.}
\label{fig:Phase_Amplitude}
\end{figure}

Recall that the eigenstate population,
\begin{eqnarray}
     p_{n} = \frac{\sin{\left[ \rchi(t_1) + E_n t_1  \right]} +  1}{\mathcal{Z}(t_1)}, 
     \label{eq:appendx_population}
\end{eqnarray}
takes a sinusoidal form after single step. In particular,
\begin{eqnarray}
     \rchi =  {\rm arg} ( \mu + i \Tilde{\mu} ) + {\rm arg} ( \mathbf{S}_{\textcolor{blue}{01}} ),
     \label{eq:phase_shift}
\end{eqnarray}
represents a phase offset determined by the Hamiltonian and overlap matrix elements, while
\begin{eqnarray}
    \mathcal{Z} = \frac{2Q \Tilde{\rho}( \Tilde{\rho} +  \rho )} {( \Tilde{\rho} + \rho )^2 \lambda_{-}^2 + g^2 \lambda_{+}^2},
    \label{eq:prop_const}
\end{eqnarray}
gives a scaling to normalize the eigenstate population. In the expressions above, ${\rm arg}(\cdot)$ denotes the argument of a complex number and $\Tilde{\mu}, \mu, \Tilde{\rho}, \rho, g, \lambda_{\pm}$ are all auxiliary variables in Eqs.~\eqref{eq:phase_shift}-\eqref{eq:prop_const}. In terms of the matrix elements, these auxiliary variables are
\begin{align}
    \lambda_{\pm} &= \frac{1}{\sqrt{ 1 \pm
    \abs{\mathbf{S}_{01}} }},
    \label{eq:auxvar_1} \\ 
    g &= \frac{\Re{\mathbf{S}_{01}} \Im{\mathbf{H}_{01}} - \Im{\mathbf{S}_{01}} \Re{\mathbf{H}_{01}} }{\abs{\mathbf{S}_{01}}} \lambda_{-} \lambda_{+}, \\
    \rho &=  \frac{\Re{\mathbf{S}_{01}} \Re{\mathbf{H}_{01}} + \Im{\mathbf{S}_{01}} \Im{\mathbf{H}_{01}} }{2\abs{\mathbf{S}_{01}}} \left( \lambda_{-}^2 + \lambda_{+}^2 \right) \\ 
    &\hspace{1cm} - \frac{\mathbf{H}_{00}}{2} \left( \lambda_{-}^2 - \lambda_{+}^2 \right), \\
    \Tilde{\rho} &= \sqrt{g^2 + \rho^2}, \\
    \mu &= \frac{2 g \lambda_{+}}{( \Tilde{\rho} + \rho ) \lambda_{-}}, \\
    \Tilde{\mu} &= \frac{- g^2 (\lambda_{-}^2 - \lambda_{+}^2) - 2 \rho (\Tilde{\rho} + \rho) \lambda_{-}^2 }{( \Tilde{\rho} + \rho )^2 \lambda_{-}^2}, 
    \label{eq:auxvar_2}
\end{align}
where $\Re$ and $\Im$ label the real and imaginary part of the matrix elements respectively. Notice that the dependence on $t_1$ is implied in the definitions of the auxiliary variables. Fig.~\ref{fig:Phase_Amplitude} shows the phase and normalization of eigenstate population for the gapped Hamiltonian as a function of the evolution time $t_1$.  Note that the phase offset $\rchi$ stays linear for a short time and then undergoes damped oscillations, where the spectral gap sets the slope and envelope in both the linear and oscillatory regimes respectively. 
The normalization factor $\mathcal{Z}$ grows quadratically for short time and then plateaus to an asymptotic value around $Q$ with intertwined oscillations, whose envelope is likewise set by the gap value. 

The condition on the phase factor $\Delta E t_1$ for derivation of Eq.~\eqref{eq:appendx_population} is as follows: at fixed $\Delta E$, invariance of $\mathbf{H}$ and $\mathbf{S}$ under periodic shift of $\Delta E t_1$ by $2\pi \mathbb{Z}$ suggests that we can always make the assumption $\Delta E t_1 \in (0, 2\pi)$. To ensure that the constituent expressions given by Eqs.~\eqref{eq:auxvar_1}-\eqref{eq:auxvar_2} are well-defined, we also exclude the measure zero subset of $t_1$ values for which $g(t_1) = 0$ and $\rho(t_1) \leq 0$. When $Q \Delta E t_1 \in 2\pi \mathbb{Z}$, it is straightforward to check that $\mathbf{S} = \mathbf{I}$ and our previous formula seems to fail with $\Re \mathbf{S}_{\textcolor{blue}{01}} = \Im \mathbf{S}_{\textcolor{blue}{01}} = \abs{\mathbf{S}_{\textcolor{blue}{01}}} = 0$. However, $\lim_{\Delta E t_1 \rightarrow 2\pi k /Q} p_n(t_1)$ exists for integer $1 \leq k \leq Q-1$ and matches with the analytical expression from diagonalization of $\mathbf{H}$. Hence Eq.~\eqref{eq:appendx_population} remains valid almost surely on the phase interval $(0, 2\pi)$.

For the case $\Delta E t_1 = \pi$ and $Q \in 2\mathbb{Z}^{+} - 1$, we can show that the extracted ground state takes a form,
\begin{eqnarray}
     | \Psi_{\rm g} (\pi/\Delta E) \rangle = \begin{cases}
     \displaystyle \sum_{1 \leq n \leq Q}^{n~{\rm odd}} \sqrt{\frac{2}{Q+1}} | n \rangle \\
     \\
     \displaystyle \sum_{1 \leq n \leq Q}^{n~{\rm even}} \sqrt{\frac{2}{Q-1}} | n \rangle \\
     \end{cases},
\end{eqnarray}
almost identical to the case $Q \in 2\mathbb{Z}^{+}$ specified within the main text, except that the solution above is degenerate. 

On the other hand, a generic choice of $\Delta E t_1 \in (0, 2\pi)$ influences the extracted population profile in a nonmonotonic way. For each eigenstate $|n\rangle$, the population $p_n(t_1)$ oscillates between its local extrema at a characteristic rate of $2\pi / [ \rchi(t_1) + n t_1 \Delta E ]$ as we vary $\Delta E t_1$. Consequently, we expect some region in the phase parameter space where the population of the low energy eigenstates fully
dominates that of the higher energy eigenstates and hence the extracted ground state $| \Psi_g \rangle$ is optimal. Such nonmonotonicty has been explicitly shown in Fig.~\ref{fig:Single_Timestep} using an optimal timestep $\Delta E t_1 \in (0, \pi/Q)$.

\section{\label{sec:Multi-step Phase Cancellation} Phase cancellation from optimized M\"{o}bius transformations}
\begin{figure}[H]
\centering
\includegraphics[width=.5\textwidth]{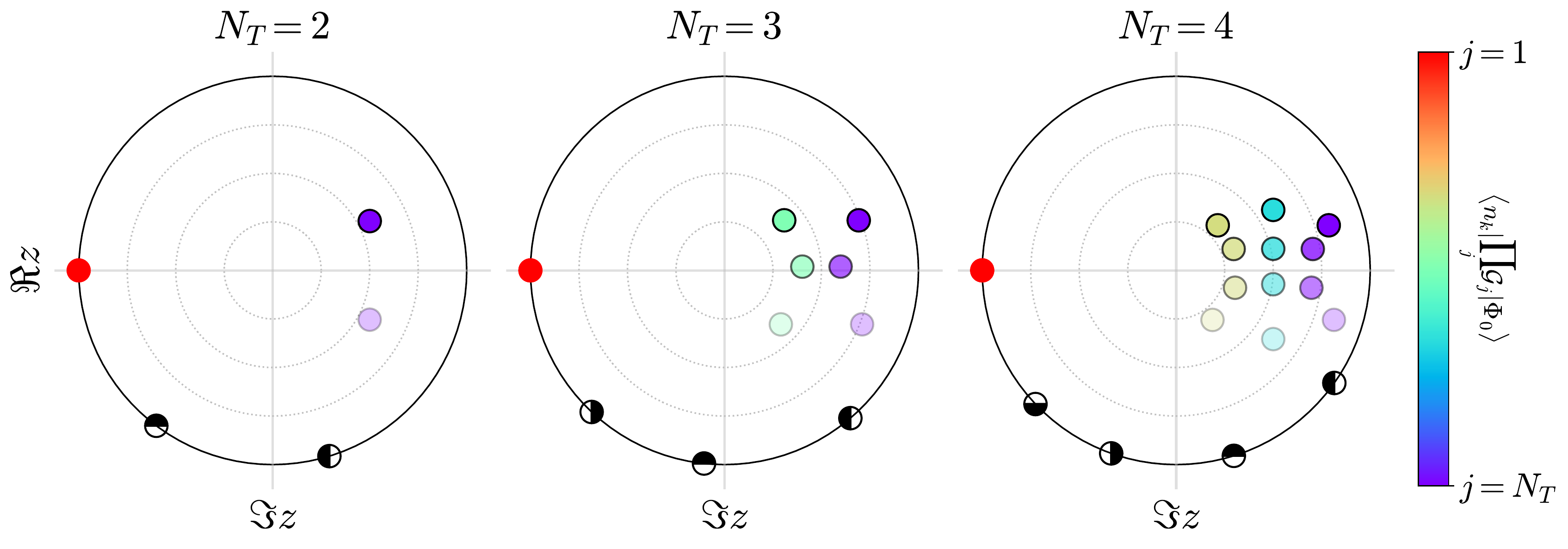}
\caption{Linear combination of time evolved states interpreted as phase rotation and tilting. The transformed variables $z_{j}(n_{k})$ are plotted as filled markers where the defining recursion in Eq.~\eqref{eq:algebraic_Phasor} is regarded as a sequence of geometric moves acting on the complex plane. Marker color highlights action of the $N_T$ moves $\mathcal{G}_{j}$ and interpolates between purple ($j=N_T$) and red ($j=1$). For any given color, marker transparency distinguishes the $N_T$ eigenstates $n_{k}$ whose population is to be suppressed after the sequence of moves. For reference, the calibrated phases $\exp{(-i E_{n_k}t_1)}$ are displayed alongside as half-filled markers on the unit circle $|z|=1$.}
\label{fig:Nested_Phasor}
\end{figure}

For multi-step time evolution, we can rewrite the extracted ground state,
\begin{eqnarray}
     |\Psi_{\rm g}\rangle \propto \prod_{j=N_T}^{1} \hat{R}_{j} | \Phi_{N_T} \rangle = \prod_{j=N_T}^{1} \hat{\mathcal{T}}_j \left[ c_j \hat{U}(\Delta t_j) | \Phi_0 \rangle \right],
     \label{eq:Lincomb_Phasor}
\end{eqnarray}
where $\hat{R}_{j}: v \mapsto |\Phi_{j-1} \rangle + c_{j} v$ defines the nested linear combinations. 
In the second equality, $\Delta t_j = t_{j} - t_{j-1}$ while $\hat{\mathcal{T}}_j$ is the $|\Phi_0\rangle$-translation of the image subspace ${\rm Im}\hat{U}(\Delta t_j)$. Eq.~\eqref{eq:Lincomb_Phasor} recapitulates that a pairwise combinator of the form $| \Phi_{j-1} \rangle + c_j | \Phi_j \rangle$ will rotate accumulated phases $\exp{(-i E_n \Delta t_{j})}$ commonly via $c_j$ (up to a stretch) and tilt the rotated phases via addition of $1$. If we project this operator identity onto eigenstate $| n \rangle$, we may view the emergent algebraic recursion $z_0(n)=1$ and,
\begin{eqnarray}
   \begin{split}
    z_{j}(n) &=
     1 + \\
     &c_{N_T-j+1} \exp{\left( - i E_n \Delta t_{N_T-j+1} \right)} z_{j-1}(n),
   \end{split}
   \label{eq:algebraic_Phasor}
\end{eqnarray}
as a sequence of M\"{o}bius transformations which direct simple geometric moves in the complex plane. In particular, each geometric move $\mathcal{G}_j$ consists of a phase calibration $z \mapsto \exp{( - i E_n \Delta t_{N_T-j+1})} z$ followed by a stretching rotation $z \mapsto c_j z$ and an additive tilt $z \mapsto z + 1$ as described above, where the last move yields the eigenstate population $|z_{N_T}(n)|^2 = |\langle n | \Psi_{\rm g}(t_1, \cdots, t_{N_T})  \rangle|^2$. Now recall that the weights $\{ c_j \}_{j=1}^{N_T}$ are optimized to maximally restrict the excited states population. Geometrically, the weights hence encode an optimal sequence of phase moves that best lower the energy cost,
\begin{eqnarray}
       \begin{split}
            \argmin_{ \{ \mathcal{G}_j \}_{j=1}^{N_T}, \{n_k\}_{k=1}^{N_\ell}: z_{N_T}(n_{k}) = 0}~& \sum_{n=1}^{Q} E_n \abs{z_{N_T}(n)}^2 \\
            &\stackrel{Q \gg 1}{\xrightarrow{\hspace*{0.8 cm}}} (c_1, \cdots, c_{N_T}),
       \end{split}
\end{eqnarray}
where the roots $\{n_k\}_{k=1}^{N_\ell}$ designate $N_\ell$ spectral landmarks of high energy cost. Heuristically, the $N_T$ complex degrees of freedom available in a move sequence are capable of handling a maximum number of $N_{\ell} = N_T$ independent nodal constraints $z_{N_T}(n_k) = 0$ so we expect a one-to-one correspondence between moves $\{ \mathcal{G}_j \}_{j=1}^{N_T}$ and distinct landmarks $\{ n_k \}_{k=1}^{N_T}$ given a suitably short evolution $\vec{t}$. For $N_T = 1$, the move $\mathcal{G}_1$ involves a supplementary rotation that locks the calibrated phase $\exp{( - i E_{n_1}t_1)} \mapsto -1$ and a subsequent counteractive tilt. For $N_T \geq 2$, a sequential move $\mathcal{G}_j$ involves a stretching rotation that clusters the calibrated phases around the negative real axis and a subsequent tilt that sends the phase cluster across the imaginary axis and thus successively flips the relative phases ${\rm arg}[z_{j}(n_k)/z_{j}(n_{k'})] \rightarrow \exp{[i (E_{n_k} - E_{n_{k'}})  t_1]}$ before the final counteraction of $\mathcal{G}_1$. These different moves are schematically shown within Fig.~\ref{fig:Nested_Phasor}. For the specific scenario $N_{\ell} = N_T = Q-1$, the set of nodal constraints may be regarded as a restriction over the energy domain dual to the phase cancellation conditions (PCCs) over the time domain explored in our previous work~\cite{Klymko2021_VQPE}.
\bigskip

\section{\label{sec:IVQPE_Preparation} Preparation of Ground State from IVQPE}
We note that the preparation of ground states in quantum circuit by IVQPE is no different than that by VQPE, even though IVQPE runs its own classical processing and generates a sequence of intermediate states as the time-evolved states. For convenience, we focus our discussion on the simple case $N_I = 1$.

\begin{figure}[tbh!]
\centering
\includegraphics[width=.45\textwidth]{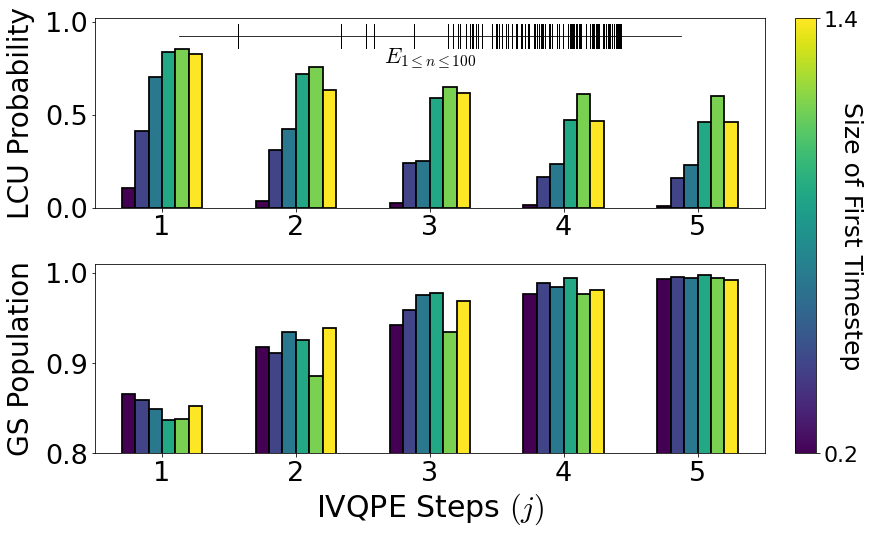}
\caption{Intermediate state preparation for IVQPE. Here we consider an adaptive time grid $\Delta t_j = 2^{j-1} t_1$ and an initial state with ground state population $0.5 < |z_1|^2 < 0.75$. \textbf{Top}: LCU success probability of the first few IVQPE steps. Spectrum of the model Hamiltonian is also shown inset. \textbf{Bottom}: Ground state population over the course of the time evolution. The color distinguishes the timestep size $\kappa = t_1(E_Q - E_1)/2\pi$, interpolating between dark green (small step size) and yellow (large step size).
}
\label{fig:LCU_prob0}
\end{figure}

In particular, our phase cancellation intuition suggests that each IVQPE step acts as a sum of two unitary gates, $\hat{I} + \exp{(i\vartheta)}\hat{U}(\Delta t_j)$, where the interfering phase $\vartheta$ changes after each iterative step. Childs and Wiebe in~\cite{LCU} provide the probability of applying linear combinations of unitary operations (LCU), thus yielding a success probability of the first iterative step,
\begin{eqnarray}
  P_{\rm success}(t_1) = \sum_{n = 1}^{Q} \abs{z_n}^2 \sin^2{\frac{
  (E_x - E_n)t_1}{2}},
\end{eqnarray}
where $E_x$ specifies the center of spectral decay. Upon an LCU measurement that indicates success, we obtain the first intermediate state $|\Phi_0\rangle \mapsto |\Psi_{\rm g}(j=1) \rangle$ with $ \abs{z_n}^2 \mapsto |z_n|^2 \sin^2{\frac{(E_x - E_n)t_1}{2}}$ (up to some overall normalization). We then proceed to implement the next sum of unitaries on this intermediate state. Hence an approximate ground state from an iterative evolution of $N_T$ timesteps can be prepared via LCU with a success probability,
\begin{eqnarray}
  P_{\rm IVQPE}(\{t_j\}) = \prod_{j = 1}^{N_T} P_{\rm success}(\Delta t_j).
\end{eqnarray}
For the probability above to take any reasonable value, the low energy amplitudes $|z_n|$ must be significant.
This can be realized in quantum chemistry applications, for example if we start with a Hartree-Fock state.

\begin{figure}[tbh!]
\centering
\includegraphics[width=.4\textwidth]{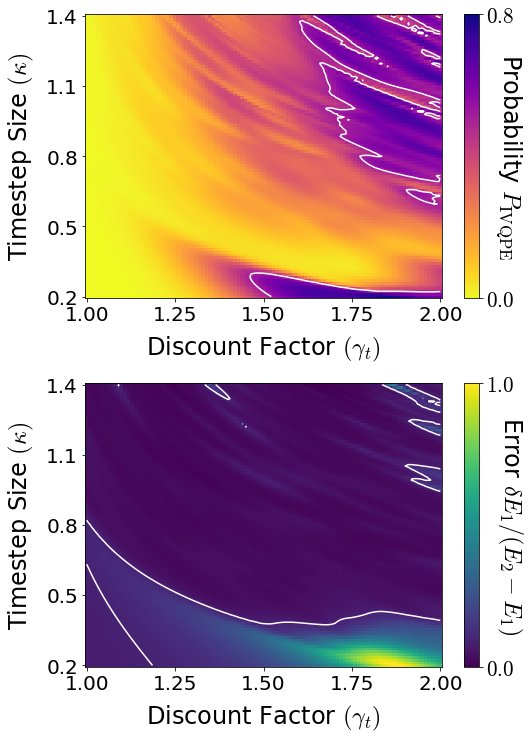}
\caption{Ground state preparation for a simulation of $N_T = 5$ timesteps (adaptive time grid $\Delta t_j = \gamma_t^{j-1} t_1$). For simplicity, we choose the same model Hamiltonian and initial state as in Fig.~\ref{fig:LCU_prob0}. \textbf{Top}: Success probability $P_{\rm IVQPE}(\{ t_j\})$ is plotted as a function of the time parameters $\kappa = t_1(E_Q-E_1)/2\pi$ and $\gamma_t$. White contour indicates the level set $P_{\rm IVQPE} = 0.5$. \textbf{Bottom}: Ground state energy error $\delta E_1 = \langle \Psi_{\rm g}(N_T) | \hat{H} | \Psi_{\rm g}(N_T) \rangle - E_1$, normalized by spectral gap, is plotted as a function of the time parameters. White contour indicates the level set $\delta E_1/(E_2 - E_1) = 0.1$.
}
\label{fig:LCU_prob}
\end{figure}

Given such an initial state, preparation of the intermediate states using a quantum circuit can be simulated and the associated probabilities of implementing the first few IVQPE steps are displayed in Fig.~\ref{fig:LCU_prob0} for a model molecular Hamiltonian. As we tune the size of the timestep, we observe individual probabilities that are relevant for practical implementation. A general time grid dependence is further investigated in Fig.~\ref{fig:LCU_prob} where, with reasonable $P_{\rm IVQPE}$, the approximate ground state can be prepared from a range of time parametrizations.

\section{\label{sec:Noise_Model}Noise Modeling from Spectral Statistics}
\begin{figure}[H]
\centering
\includegraphics[width=.45\textwidth]{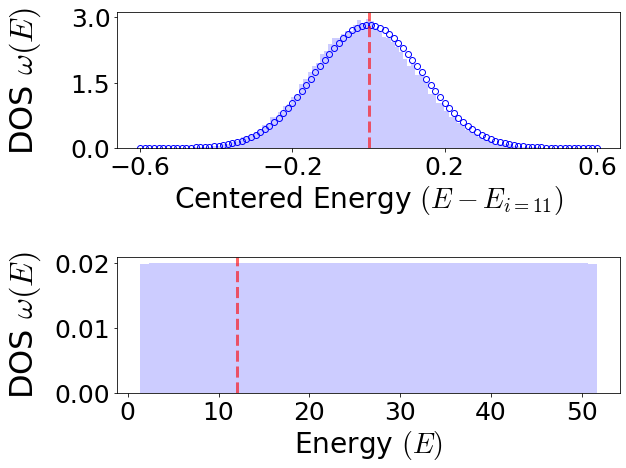}
\caption{Broadening of the spectral DOS from the phenomenological noise model. 
We consider an unperturbed Hamiltonian with linear spectrum $E_n = n \Delta E$ and a Gaussian perturbation $||\hat{V}(t)|| \approx \Delta E$. \textbf{Top}: DOS of the perturbed Hamiltonian close to a specific unperturbed energy eigenvalue. Analytical prediction of $g_n(E)$ from perturbation theory is displayed as blue circles. \textbf{Bottom}: Global DOS of the perturbed Hamiltonian. Histograms are computed from $5 \times 10^4$ noise realizations and the red dashed line marks the unperturbed energy $E_{11}$.
}
\label{fig:Broaden}
\end{figure}

To see how the target spectral DOS $\omega(E)$ can broaden in the presence of noise, let us consider a phenomenological model~\cite{Noise_2005,Noise_2019} for which the Hamiltonian $\hat{H}$ undergoes a Hermitian stochastic perturbation $\hat{H} \mapsto \hat{H} + \hat{V}(t)$ during the time evolution. $\hat{V}(t)$ in the computational basis is taken to be a Gaussian random matrix, 
\begin{equation}
  \hat{V} \sim \frac{\exp{\big[ -Q{\rm tr}(\hat{V}^2)/4 \big]}}{2^{Q/2} (2\pi)^{Q(Q+1)/4}},
  \label{eq:GOE}
\end{equation}
and for now we assume a memoryless perturbation, \textit{i.e.}, $\hat{V}(t)$ is uncorrelated with $\hat{V}(t')$ unless $t = t'$. Without loss of generality, we are free to make a change basis since Eq.~\eqref{eq:GOE} is invariant under any similarity transformation. Thus in the eigenbasis of $\hat{H}$, we have
\begin{equation}
 \begin{split}
      E_n \mapsto & E_n \\
      & + \langle n | \hat{V} | n \rangle + \sum_{m \neq n} \frac{ | \langle n | \hat{V} | m \rangle|^2 }{E_n - E_m} + O\big( ||\hat{V}||^3 \big),
 \end{split}
\end{equation}
using standard results from perturbation theory in quantum mechanics. Up to first order (in the operator norm of the perturbation), we notice that the DOS becomes
\begin{equation}
    \omega(E) = \sum_{n=1}^{Q} \delta(E-E_n) \mapsto \sum_{n=1}^{Q} g_n(E),
\end{equation}
where the broadening $g_n$ is given by a Gaussian centered at $E_n$. Similarly, higher order corrections leads to a distinct functional form of $g_n$ as long as we remain in the perturbative regime. 
Such spectral broadening is illustrated in Fig.~\ref{fig:Broaden}

\bibliographystyle{unsrtnat}
\bibliography{mainbib,katie}
\end{document}